 \documentclass[prb,aps,showpacs,a4paper]{revtex4}
\usepackage{dcolumn,epsfig}
\begin{document}
\title{Density functional simulation of small Fe nanoparticles}
\author{A.~V.~Postnikov}
\email{postnik@thp.uni-duisburg.de}
\author{P.~Entel}
\email{entel@thp.uni-duisburg.de}
\affiliation{Gerhard Mercator University Duisburg -- 
Theoretical Low-Temperature Physics,
D-47048 Duisburg, Germany}
\author{Jos\'e M. Soler}
\email{jose.soler@uam.es}
\affiliation{Departamento de F\'{\i}sica de la Materia Condensada,
Universidad Aut\'onoma de Madrid, E-28049 Madrid, Spain}
\date{\today}
\begin{abstract}
We calculate from first principles the electronic structure, 
relaxation and magnetic moments of small Fe particles,
by applying the numerical local orbitals method in combination
with norm-conserving pseudopotentials. The accuracy of
the method in describing elastic properties and magnetic
phase diagrams is tested by comparing benchmark results
for different phases of crystalline iron to those obtained
by an all-electron method. Our calculations for the
bipyramidal Fe$_5$ cluster 
confirm previous plane-wave results that predicted
a non-collinear magnetic structure.
For larger bcc-related (Fe$_{35}$, Fe$_{59}$) and fcc-related
(Fe$_{38}$, Fe$_{43}$, Fe$_{55}$, Fe$_{62}$) particles, 
a larger inward relaxation of outer shells has been found
in all cases, accompanied by an increase of local magnetic
moments on the surface to beyond 3 $\mu_{\mbox{\tiny B}}$.
\end{abstract}
\maketitle
\section{Introduction}
Magnetic properties of small iron nanoparticles are often unusual and
quite different from those of the bulk. 
Experimental information is not abundant and confined essentially
to determination of mean magnetic moment, depending on the cluster
size and temperature, e.g. from Stern-Gerlach 
deflection as measured by Billas \emph{et al.} \cite{PRL71-4067,Sci265-1682} 
(see also Ref.~\onlinecite{PRB62-7491} for a recent review on experimental 
situation). The morphology, structural relaxation or distribution of magnetic
moments from the center to the surface are so far not accessible
in experiment. The microscopic theory, on the other hand,
can in principle address these issues and thus be of great help.
A range of different approaches have been tried on Fe clusters,
illustrating a compromise between severeness of approximations done and
the size of system to be treated. The lack of symmetry is
a common difficulty in all cluster studies; but Fe poses an
additional challenge even among other transition-metal systems
because its electronic structure and magnetic properties are known to be 
very sensitive to local environment.
In course of the last decade, a number of simulations have been done
with the use of appropriately tuned potential energy 
functions (see, e.g., Christensen and Cohen\cite{PRB47-13643}
and Besley \emph{et al.} \cite{JMC341-75}),
or of parametrized Hubbard-type Hamiltonian, to account for
magnetic properties \cite{PRB40-7642,PRB53-10382,ChPL260-15,EPL36-37,%
PRB57-10069,PRB54-3003}. The parameterization is typically done
to the data of (bulk) band structure calculation and experimental
(spectroscopic) results; common approximations are the neglection of
interactions beyond the first (or second) neighbors, sometimes
fixed cluster geometry \cite{PRB55-13283} or only topological
variation of unrelaxed cluster structure, with nearest-neighbors distances
fixed \cite{PRB53-10382}. Those semiempirical calculations
which keep track of electronic degrees of freedom and not just
potential function were able to treat up to 89 
(Ref.~\onlinecite{PRB54-3003}) --
169 (Ref.~\onlinecite{PRB57-10069}) -- 
177 (Ref.~\onlinecite{PRB55-13283}, in a fixed high-symmetry arrangement)
Fe atoms.

First-principles simulations by means of density functional 
theory (DFT) \cite{RMP71-1253}, 
while still subject to certain basic approximations, 
have the advantage of not being biased by a particular parameterization.
Following constrained structure optimizations (for selected
high-symmetry arrangements) of small (up to Fe$_4$) systems
by Chen \emph{et al.} \cite{PRB44-6558}, Castro and Salahub
searched for ground-state structures, and reported other
related properties, of low-symmetry clusters, Fe$_5$ 
being the largest\cite{PRB49-11842,ChPL271-133}.
Recently Kortus \emph{et al.} \cite{APL80-4193}
calculated magnetic moments and anisotropy energies
of (symmetric) 5-at. and 13-at. clusters of Fe-Co composition,
including all-Fe case.
Ballone and Jones \cite{ChPL233-632} optimized the structures
of clusters up to Fe$_7$, making use of norm-conserving
pseudopotentials and Car-Parrinello molecular dynamics.
A similar approach, i.e. essentially a planewave simulation
of a cluster in a box with implied periodic boundary conditions,
has been applied in later publications by Oda \emph{et al.} \cite{PRL80-3622}
and Hobbs \emph{et al.} \cite{PRB62-11556} who concentrated on apparent
non-collinearity of local magnetic moments
in the ground-state structures of Fe$_3$ and Fe$_5$ clusters.
The results of these two calculations, while being not quite identical, 
contested previous results obtained for the cluster geometry and energetics
obtained by a (magnetically collinear) methods
(e.g., by Gaussian orbitals technique in Ref.~\onlinecite{ChPL271-133})
and seem to establish a present-day margin of reliability for 
DFT-based calculations. A snapshot of available theoretical
knowledge on transition-metal clusters as for 1999, along with
relevant experimental information, can be found in a review by
Alonso \cite{ChemRev100-637}.

While for such relatively small systems one can hope to
pinpoint the ground-state geometry among several competing
metastable configurations, this seems hardly feasible for
larger particles, consisting of several tens or hundreds of atoms.
However, in such systems some general ideas about the particles'
morphology, radial distribution of density and magnetization
would be already of interest for establishing relation with
measurements on particle beams, where clusters of such sizes
typically participate and exhibit non-trivial variation of properties
with size \cite{PRL71-4067,Sci265-1682}.
It would be a reasonable assumption that bulk-like features,
i.e. at least local neighborhood of either bcc or fcc type
would emerge with increasing the particle size. Moreover, 
the arrangement of atoms over icosahedral shells has been detected
for rare gas clusters and also comes into consideration
for Ni particles (see Ref.~\onlinecite{ChemRev100-637} for a review).
One should note that a self-consistent treatment of electronic
structure over a Fe particle including several shells of atoms
is a technically demanding task even for a fixed atomic
arrangement, and the need for accurate total energy and forces
for a structure optimization complicates this problem even further.
Fujima and Yamaguchi \cite{MSE217-295} performed DFT calculations 
by a discrete variational method for Fe$_{15}$ and Fe$_{35}$ clusters 
in the fixed (crystalline-like) bcc structure,
that exhibited an enhancement of local magnetic
moments towards the surface. 
Recently Duan and Zheng \cite{PhLA280-333} reported magnetic
moments in 13- and 55-atom clusters of Fe, Co and Ni 
in fcc-, hexagonal closed packed and icosahedral geometry,
allowing a breathing relaxation within the DFT (for small clusters only).

The aim of the present paper is to study electronic structure
and magnetism in some additional representative 
cluster morphologies, making use of the first-principles method
within the DFT that would allow accurate evaluation of total energy
differences, conjugate gradient optimization of ground-state
structure without need for symmetry constraints. To this end, we apply
{\sc Siesta}, \cite{IJQC65-453,PSSB215-809,JPCM14-2745} an \emph{ab initio}
electronic-structure and molecular-dynamics simulation package,
relying on norm-conserving pseudopotentials and strictly localized 
numerical pseudoatomic orbitals.
This method has recently been applied to the relaxation of gold clusters, 
up to Au$_{75}$ \cite{PRB61-5771}.
Also, some studies have been done by Izquierdo \emph{et al.} \cite{PRB61-13639} 
on bulk and low-dimensional Fe systems (small free and deposited clusters,
monolayer and nanowire) \cite{PRB61-13639} -- primarily, in the view of
their magnetic properties and without structure relaxation allowed.
Finally, Di{\'e}guez \emph{et al.} \cite{PRB63-205407} searched for 
ground-state structures and evaluated the average magnetic moment and 
other properties in the hierarchy of Fe clusters from 2 to 17 atoms.

In our simulation of larger clusters we want to be sure that
the accuracy of the {\sc Siesta} method is sufficient for the adequate
description of delicate structure/magnetization interplay in Fe systems.
Therefore, after specifying the details of calculation
we report the benchmark calculations for bcc/fcc
phase diagram of bulk iron. Further on, we allow for non-collinear
alignment of magnetic density in the treatment of the Fe$_5$
cluster as another benchmark, that is discussed afterwards
in relation with previous planewave calculations on this system.
In the subsequent sections, we present the results for nanoparticles 
(in bcc and fcc prototype structure, and with the icosahedral symmetry) 
not covered by previous studies, with the 62 Fe atoms as the largest.

\section{Computational procedure}
The underlying methodology and performance of the {\sc Siesta} method 
has been reviewed in recent publications\cite{PRB51-1456,PRB53-10441,%
JPCM14-2745}. The basis set consists of numerically defined
strictly localized pseudoatomic orbitals. 
It uses norm-conserving (possibly hard) pseudopotentials and
a uniform real-space grid to represent the valence charge density
and to calculate the Hartree and exchange-correlation potentials
and their corresponding matrix elements.

The necessary plane-wave cutoff in this charge density representation
may become large, depending on the system under consideration.
In many systems, e.g. those containing magnetic $3d$ elements, 
it turns out essential to use a pseudocore correction for the pseudopotential 
as proposed by Louie \emph{et al.} \cite{PRB26-1738}, in order
to accurately recover the exchange-correlation potential due to
full density, imitating the effect of the core states.
A strong localization of such pseudocore for Fe demands
a high charge-density cutoff (see below), that is
one of the main limiting factors in the treatment of larger Fe systems
(in what regards the size of the simulation cell) by {\sc Siesta}.
The extension of the most diffuse basis function on Fe atom was 
about 6.8 a.u., and the resulting size of the simulation cell
for a cluster (i.e. that would prevent direct overlap
of atomic functions across the cell boundary)
was 35 a.u. for the largest particle we used (Fe$_{62}$).

We used a double-$\zeta$ singly-polarized (DZP) basis set, with 15
orbitals per atom. This basis was discussed for Fe 
in Ref.~\onlinecite{PRB61-13639}).
The pseudopotential has been generated for the [Ar]$3d^7\,4s^1$
configuration (with $3p^6$ sometimes also participating in the
pseudopotential generation, see below) and a pseudocore radius of 0.7 a.u.
Other aspects of calculation were largely similar
to those of Ref.~\onlinecite{PRB61-13639}.
We note that the same calculation method but slightly
different setup has been recently used 
by Di\'eguez \emph{et al.} \cite{PRB63-205407}.
The most important difference is the use of triple-$\zeta$
basis set with double-$\zeta$ polarized functions there, which is
probably superior to our choice of a more compact basis.
Moreover, the spatial extension of the basis functions 
in the calculation by Di\'eguez \emph{et al.}
were larger \cite{note_on_basis}.
The latter fact lead to a considerably larger size
of the simulation cell and, with the cutoff
chosen in Ref.~\onlinecite{PRB63-205407}, to a relatively sparse mesh
for the representation of the charge density. 
However, we used a higher cutoff and hence more dense real-space mesh.

A part of the results of the present study deals with
a general-shape (non-collinear) treatment
of magnetization. Whereas its implementation in the
local density approximation (LDA) may be considered as
relatively straightforward, with the use of a
generalized gradient approximation (GGA) the exchange-correlation
energy in non-collinear case would depend on the gradients of both the
magnitude and the direction of the magnetization. To our knowledge,
no such full implementation has been elaborated, and the
non-collinear GGA calculation of Hobbs \emph{et al.} \cite{PRB62-11556}
was not sufficiently detailed on this point. Thus, we restrict the
non-collinear calculations to LDA.

\section{Benchmark calculations for bulk iron}
\begin{figure}[p]
\centerline{\epsfig{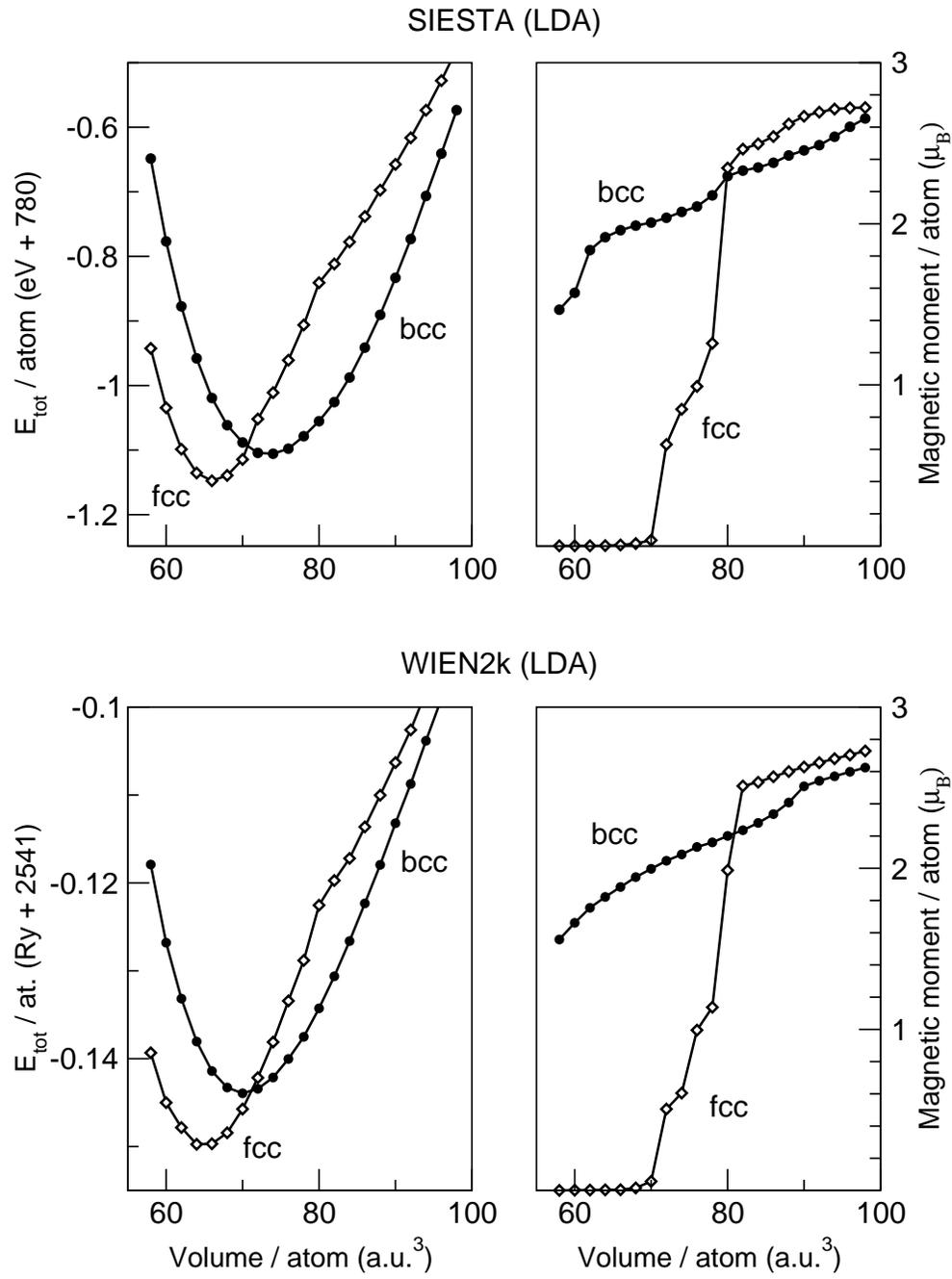}}
\bigskip
\caption{Total energy vs. volume (left panel)
and magnetic moment per atom (right panel) for fcc and bcc phases
as calculated with {\sc Siesta} (top) and WIEN2k (bottom)
in the local density approximation.
Brillouin zone integration in {\sc Siesta} was done by the
Monkhorst-Pack method with 12$\times$12$\times$12 divisions
in the full Brillouin zone.
}
\label{fig:LDA}
\end{figure}

The electronic structure of bulk iron is well known and
can nowadays be well reproduced by different methods
of density functional theory.
The {\sc Siesta} method was earlier used 
by Izquierdo \emph{et al.} \cite{PRB61-13639}
to study the ground state
properties (equilibrium volume, bulk modulus, magnetic moment)
of bcc-Fe using the GGA.
A comparison with results obtained by other methods can also be found
in this publication. However, two important issues of the ground state
of Fe as a benchmark system of the DFT were not addressed 
in Ref.~\onlinecite{PRB61-13639} and are discussed below.
These are, a low-spin to high-spin transition with the increase
of volume in fcc-Fe \cite{PRB34-1784,PRL57-2211}, 
and the necessity to go beyond the LDA
in order to obtain a correct sequence of bcc--fcc ground state
energies \cite{PRB40-1997,PRB44-2923}.
The calculated energy--volume curves for different structural
and magnetic phases of Fe can be found in Ref.~\onlinecite{PRB60-3839}.
In order to check the quality of pseudopotential and
basis set used in the present simulation, we compared the
results obtained for bulk iron using the {\sc Siesta} code
with those produced by the full-potential augmented
plane-wave method, as implemented
in the WIEN2k package \cite{wien2k}. The energy--volume curves
and magnetic moment values as calculated in the LDA for bcc
and fcc ferromagnetic phases are shown in Fig.~\ref{fig:LDA}.

It is noteworthy that the curvatures of both bcc and fcc 
energy-volume functions, not only near their corresponding minima
but over whole range of relevant volumes, are faithfully reproduced
by {\sc Siesta}. In particular, we emphasize the low spin --  high spin 
transition in the fcc phase (probably, reported for the first time
by Moruzzi \emph{et al.} \cite{PRB34-1784})
with a related kink in the energy -- volume curve.
Moreover, {\sc Siesta} accurately describes a crossover between 
the bcc and fcc volume curves and the (erroneous) result that the absolute 
energy minima occurs in the fcc phase, as is well known
to be an artifact of the LDA.
As regards the magnetic moments, it is encouraging that their
absolute values (per atom), as calculated by {\sc Siesta} and WIEN2k, 
agree well in the bcc phase over all relevant range of volumes.
In the fcc phase, the position, the magnitude and even a complicated
profile of the low spin -- high spin transition is correctly reproduced.

\begin{figure}[p]
\centerline{\epsfig{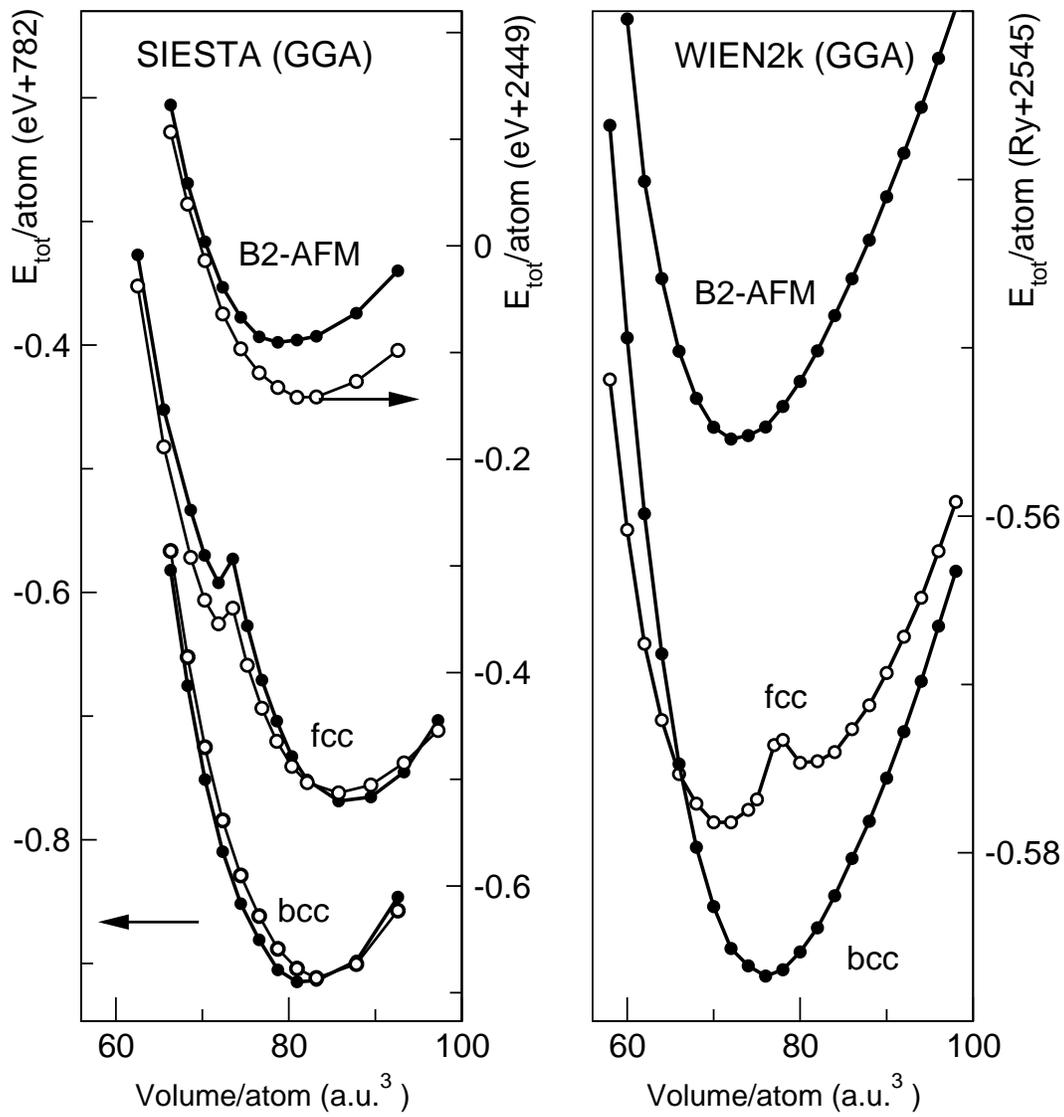}}
\bigskip
\caption{Total energy vs. volume for ferromagnetic fcc and bcc and
for the B2 antiferromagnetic phase, using generalized
gradient approximation for exchange-correlation. Left panel: {\sc Siesta}
calculation, with $3p$ states attributed to the core
(black dots, left energy scale) or treated as valence  
states in the generation of norm-conserving pseudopotential
(open dots, right energy scale). Right panel: all-electron
calculation with the WIEN2k code.}
\label{fig:GGA}
\end{figure}

A similar calculation, that also included the antiferromagnetic
(AFM; B2-type) phase for bcc iron, has been performed with the
use of GGA (Perdew-Bourke-Ernzrhof \cite{PRL77-3865} in both
{\sc Siesta} and WIEN2k realizations); the results are shown
in Fig.~\ref{fig:GGA}. The energy differences between
phases at their corresponding equilibrium volumina are fairly well
reproduced by {\sc Siesta}; the relative depths of two local energy
minima on the fcc energy curve come out however somehow distorted. 

In order to study the effect of inclusion of upper core states (Fe$3p$)
in the pseudopotential generation on the equation of state, we calculated
the total energy as function of volume with these states attributed
either to the core or to the valence band. As can be seen in
Fig.~\ref{fig:GGA}, the differences are noticeable for the
spin-flip transition energy but almost negligible for the comparison
of bcc and fcc phases. Since no equilibrium volume
nor the compressibility are affected, we won't expect 
a big role of the Fe$3p$ states included beyond the fixed core
on the structure relaxation or lattice dynamics 
with a fixed magnetic ordering.

\section{Results for the F\lowercase{e}$_5$ cluster}
The geometry and magnetic structure of small Fe clusters
(Fe$_2$ -- Fe$_5$) have been recently addressed 
in a number of \emph{ab initio}
calculations by different methods. For Fe$_3$ and Fe$_5$,
a non-collinear magnetic ordering has been
reported \cite{PRL80-3622,PRB62-11556}.
Although direct experimental
justification of these theoretical prediction was
apparently not yet possible, a qualitative consensus
between recent high-accuracy data allows to consider
these small clusters as non-trivial benchmark systems
(it should be noted that spin-orbit interaction was not considered
in the simulations cited, nor did we include it in our present
treatment). For the linear Fe$_3$ cluster, we
obtained the results in good agreement with 
the calculation of Oda \emph{et al.} \cite{PRL80-3622}: 
interatomic distance of 3.63 a.u.; 
magnetic moments of 3.07 $\mu_{\mbox{\tiny B}}$ 
(almost antiparallel but canted each by 5$^{\circ}$) at apical Fe atoms;
1.26 $\mu_{\mbox{\tiny B}}$ (coplanar with these two
and canted by 90$^{\circ}$) on the central atom.
It should be noted that Hobbs \emph{et al.} \cite{PRB62-11556}
claim this magnetic structure to be realistic but not fully converged
and unstable with respect to a fully collinear configuration.
So additional studies are probably needed to clarify
this controversy. 

%
%
\begin{table}[b]
\caption{Structure and magnetic properties of the Fe$_5$ cluster
(trigonal bipyramide) from {\sc Siesta} calculations
and from the literature. Subscript b refers to basal Fe atoms,
a -- to apical ones. Interatomic distances $a_{\mbox{\tiny b-b}}$,
$a_{\mbox{\tiny b-a}}$ are in {\AA}, 
magnetic moments $M$ in $\mu_{\mbox{\tiny B}}$.
$\Delta E$ (meV) is difference in binding energy per atom between
ferromagnetic and non-collinear configurations.}
\begin{ruledtabular}
\begin{tabular}{lddddddc}
 & \multicolumn{1}{c}{$a_{\mbox{\tiny b-b}}$}        & 
   \multicolumn{1}{c}{$a_{\mbox{\tiny b-a}}$}        &
   \multicolumn{1}{c}{$M_{\mbox{\tiny b}}$}          & 
   \multicolumn{1}{c}{$M_{\mbox{\tiny a}}$}          &  
   \multicolumn{1}{c}{canting $M_{\mbox{\tiny a.}}$} & 
   \multicolumn{1}{c}{$M_{\mbox{\tiny tot}}$}        &  
   $\Delta E$ \\
\hline
\multicolumn{8}{c}{present calculation} \\
coll. GGA (FM)  & 2.46 & 2.38 & 3.64 & 3.54 &
 \multicolumn{1}{c}{$-$}&18.00\\
coll. LDA (FM)  & 2.36 & 2.31 & 3.63 & 3.56 &
 \multicolumn{1}{c}{$-$}&18.00\\
 non-coll. LDA  & 2.34 & 2.27 & 3.40 & 3.32 &
 \multicolumn{1}{c}{40.6$^{\circ}$} & 15.24 & 25\\
\multicolumn{8}{c}{Oda \emph{et al.}\protect\cite{PRL80-3622}} \\
coll. LDA (FM)  & 2.37 & 2.22 & 2.58 & 2.55 &
 \multicolumn{1}{c}{$-$}&14.00\\
non-coll. LDA   & 2.34 & 2.25 & 2.72 & 2.71 &
 \multicolumn{1}{c}{29.7$^{\circ}$} & 14.57 & 10\\
\multicolumn{8}{c}{Hobbs \emph{et al.}\protect\cite{PRB62-11556}} \\
coll. GGA (FM)  & 2.39 & 2.34 & 3.11 & 3.17 &
 \multicolumn{1}{c}{$-$}&18.00\\
non-coll. GGA   & 2.38 & 2.33 & 3.04 & 2.71 &
 \multicolumn{1}{c}{31.3$^{\circ}$} & 15.9  & 14\\
coll. LDA (FM)  & 2.34 & 2.24 & 2.80 & 2.85 &
 \multicolumn{1}{c}{$-$}&14.00\\
non-coll. LDA   & 2.33 & 2.24 & 2.87 & 2.83 &
 \multicolumn{1}{c}{35.9$^{\circ}$} & 14.5  & 32\\
\end{tabular}
\end{ruledtabular}
\label{tab:Fe5}
\end{table}

As for the Fe$_5$ cluster, both previous studies agree at least qualitatively 
on the ground-state magnetic configuration. In Table \ref{tab:Fe5}, we 
summarize our results in comparison with those of 
Refs.~\onlinecite{PRL80-3622,PRB62-11556} for collinear 
and non-collinear arrangement 
of local moments. In our calculations, the simulation cell of the dimensions
12$\times$12$\times$12 {\AA}$^3$ was used, large enough to prevent the overlap 
of localized basis functions. The charge density was expanded in a grid of
150$\times$150$\times$150 points, corresponding to a plane-wave cutoff of 
430 Ry, and the Hartree potential was obtained by fast Fourier transformation.

\begin{figure}[p]
\centerline{\epsfig{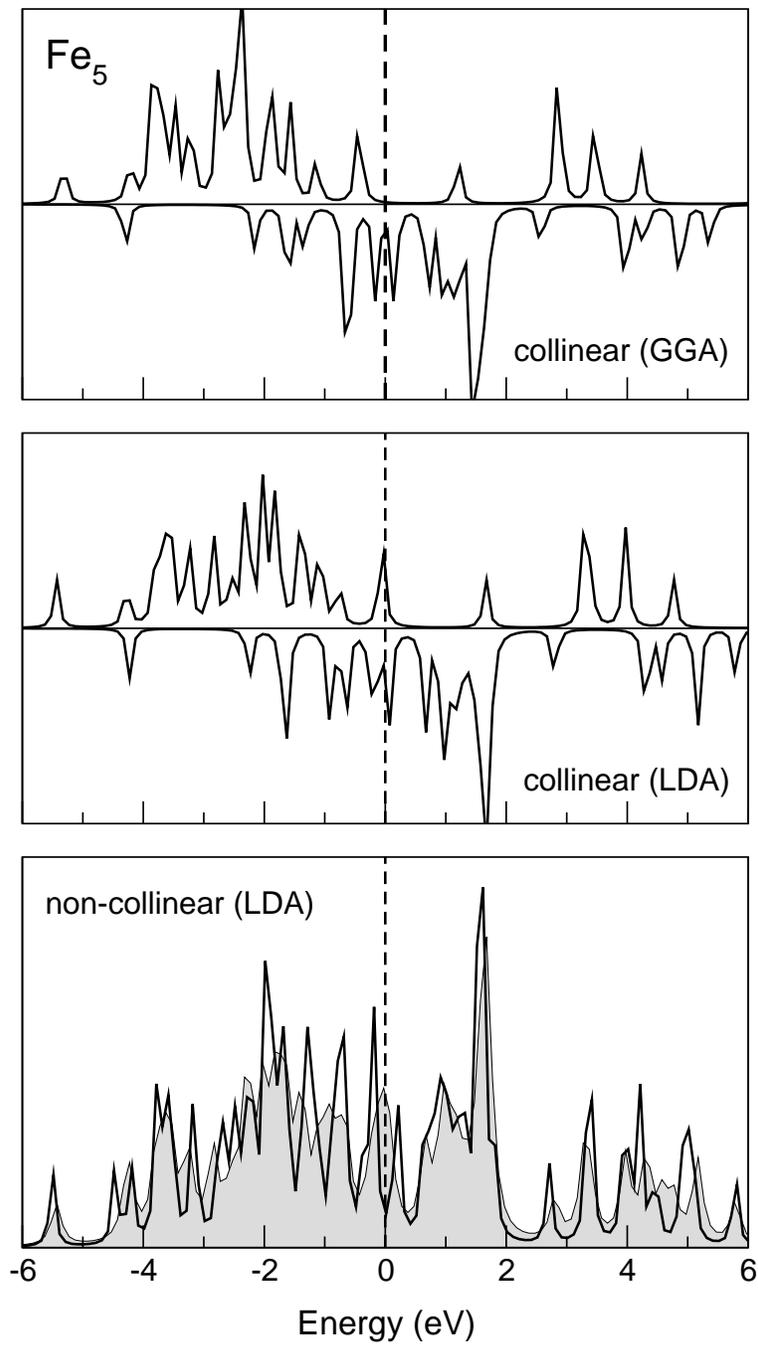}}
\bigskip
\caption{Energy levels broadened by 0.1 eV in the Fe$_5$ cluster
based on LDA and GGA calculations for ferromagnetic ordering
(top and middle panels), and for the ground-state non-collinear structure 
(bottom panel). The results for the collinear case are resolved in 
majority-spin and minority-spin contributions.
In the bottom panel, the collinear LDA result summed up over both spin channels
is shown for comparison as shadowed area. Note the elevated density of states 
at the Fermi energy (=0 in the plot), removed by non-collinear spin arrangement.
}
\label{fig:Fe5}
\end{figure}

In our present study as well as in those by Oda \emph{et al.} and 
Hobbs \emph{et al.}, a calculation constrained to a unique magnetization 
direction, i.e. a collinear one, unavoidably resulted in a stable ferromagnetic
solution. Relaxing the collinearity condition, again consistently with 
previous calculations, reliably reproduces a canted (in opposite directions)
configuration of magnetic moments on two apical atoms, with respect to basal 
atoms whose moments remain parallel. The energy gain due to forming
a canted structure lies in between the estimations done with the LDA in 
Refs.~\onlinecite{PRL80-3622} and \onlinecite{PRB62-11556}, closer to 
the latter. The values of local magnetic moments of all five atoms are very 
close, with the basal moments slightly higher in all cases of non-collinear 
structures considered. The absolute values of local magnetic moments depend on
definition: in Ref.~\onlinecite{PRB62-11556}, the magnetization density
was projected onto a sphere with radius 1.2 {\AA}, whereas in our calculation, 
the Mulliken population analysis has been done for (strictly localized) basis 
functions. The total magnetic moment obtained in our calculation with the GGA
is 18 $\mu_{\mbox{\tiny B}}$, consistently with the result of 
Hobbs \emph{et al.} \cite{PRB62-11556}. The same total magnetic moment was 
obtained in the LDA; the equilibrium interatomic distances were however
noticeably reduced as compared to the GGA (Table \ref{tab:Fe5}).
The diagram of broadened energy levels (Fig.~\ref{fig:Fe5}) shows 
that an accidental near degeneracy of two states with opposite spin direction
occurs at the Fermi level, but is removed both in the GGA and by allowing 
non-collinear spin arrangement in the LDA. The differences in the energy level 
structure from the LDA and GGA calculations seen in Fig.~\ref{fig:Fe5} are 
in part due to differences in relaxed geometry and in part due to
the exchange-correlation potential as such. GGA tends to produce larger 
separation between centers of gravity of majority-spin and minority-spin 
states than LDA does. The effect of geometry is primarily manifested in
rearranging the levels in the vicinity of the Fermi level, aimed at reducing 
the band energy.

It is difficult to estimate \emph{a priori} which part of
existing differences from earlier calculations has to do with certain 
limitations of {\sc Siesta}, like the compactness of its localized
basis set, and what can be due to other technical differences,
like the construction and treatment of pseudopotentials.
From one side, planewave methods allow a systematic enhancement
of the basis set completeness; on the other side, the use of pseudopotential
presumes certain compromise between its softness and transferability,
not trivial in relation to transition metals, or anyway ambiguities 
in its construction (even in case of ultrasoft 
pseudopotentials\cite{note_on_PAW}).
We recall that the use of {\sc Siesta}, contrary to
planewave methods, does not necessarily presume the pseudopotential 
to be soft -- although that is normally advantageous.

\begin{figure}[p]
\centerline{\epsfig{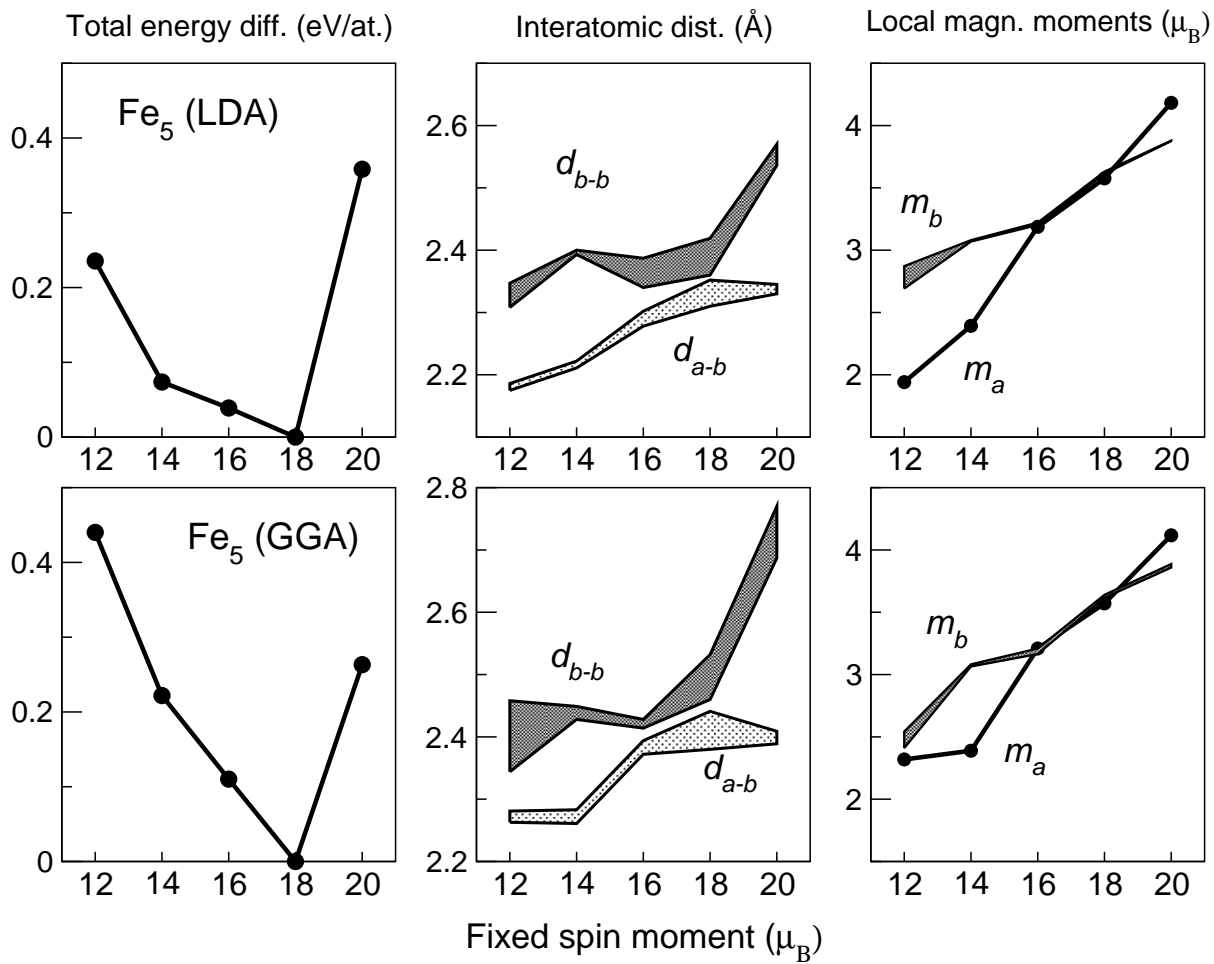}}
\bigskip
\caption{Total energy (relative to the ground state),
local magnetic moments $M_b$, $M_a$ 
and interatomic distances $d_{\mbox{\tiny b-b}}$,
$d_{\mbox{\tiny a-b}}$ depending on the fixed spin moment
as results of the LDA (upper row) and GGA (lower row)
calculations for the Fe$_5$ cluster.
Subscript \emph{b} refers to basal Fe atoms,
\emph{a} -- to apical ones.
Shaded areas indicate variation of $d$ or $m$ within different
atoms of the same type (\emph{a} or \emph{b}).
}
\label{fig:FSM}
\end{figure}

In order to bring in some systematics
in the analysis of our LDA and GGA results, we performed a sequence
of fixed spin moment calculations, 12 to 20 $\mu_{\mbox{\tiny B}}$, 
everywhere allowing for full structural relaxation.
While the fixed spin moment of 18 $\mu_{\mbox{\tiny B}}$
yielded the ground state in both LDA and GGA cases, the total
energy of the (next) $M$=16 $\mu_{\mbox{\tiny B}}$ state
is higher by only 38 meV/at. in the LDA calculation (110 meV/at. in the GGA).
Actually, even the next-stable structure with the fixed spin moment of
14 $\mu_{\mbox{\tiny B}}$ is not so much higher in energy
than the ground state (74 meV/at.). Therefore the fact that it
materialized as the LDA ground state in two previous
planewave calculations \cite{PRL80-3622,PRB62-11556}
seems plausible, in view of technical differences in implementing the LDA.
On the contrary, the GGA solution with the total moment of
18 $\mu_{\mbox{\tiny B}}$ corresponds to a deep energy minimum,
unambiguously found in both our calculation and that by Hobbs \emph{et al.}
\cite{PRB62-11556}.
The  variation of local magnetic moments and interatomic distances
over different fixed-moment states is shown in Fig.~\ref{fig:FSM}. 
First we note that with the total magnetic moment forced to be
14 $\mu_{\mbox{\tiny B}}$ in the LDA calculation, our relaxed
interatomic distances are indeed very close to those of 
Ref.~\onlinecite{PRL80-3622,PRB62-11556}.
Moreover it is instructive to discuss the evolution of 
the structure and of local magnetic moments depending on the total
fixed moment. It is understandable that larger interatomic distances 
are needed, on the average, to support larger total moments. However,
basal-basal and basal-apical distances grow with the magnitude
of the fixed spin moment at different rate.
First the apical atoms move away from a roughly fixed
triangular base (whereby their local magnetic moments get
a sharp increase); near 18 $\mu_{\mbox{\tiny B}}$
the distance between apical and basal atoms gets stabilized, and the
basal triangle grows further, as the main effect
(in both LDA and GGA cases).

A more detailed insight into the composition of hybridized electronic states 
in cluster reveals the following. As the (fixed) total spin moment $M$ 
augments, the majority-spin occupation numbers steadily grow, 
maintaining on all atoms roughly the same magnitude. 
Their minority-spin counterparts behave differently -- 
some of them drop down by 0.15--0.2 $e$, when corresponding hybridized 
states float upwards of the Fermi level. 
Thus, from $M$=14 to 16 $\mu_{\mbox{\tiny B}}$ 
the $d_{x^2-y^2}$ occupation at the apical atoms drops simultaneously
with the $d_{xz}$ of basal atoms. This is accompanied  by the increase of
the apical-basal bond length. Further on, between $M$=18 and 20 
$\mu_{\mbox{\tiny B}}$ the minority-spin $d_{z^2}$ occupation
of the basal atoms drops down, that loosens the bonding
within the basal triangle (see Fig.~\ref{fig:FSM}).
For $M\!<\!16$ $\mu_{\mbox{\tiny B}}$ and 
$M\!>\!18$ $\mu_{\mbox{\tiny B}}$, local magnetic moments of apical atoms 
grow faster than those of basal atoms. It is noteworthy that the ground-state 
structure materializes when both local magnetic moments, on one side, and
nearest-neighbor distances, on the other side, become ``balanced''
over the cluster.

In a recent study, Kortus \emph{et al.} \cite{APL80-4193} calculated 
equilibrium structure and magnetic properties of the Fe$_5$ cluster and found 
the ground-state magnetic moment of 16 $\mu_{\mbox{\tiny B}}$ (with the GGA).
They attribute the difference from our present results (and those by 
Hobbs \cite{PRB62-11556}) to the use of pseudopotentials, emphasizing 
at the same time good agreement of their data with earlier all-electron 
calculation by Castro \emph{et al.} \cite{ChPL271-133}. However, in view 
of the closeness of computational schemes applied in these two all-electron 
calculations (Gaussian-type orbitals as basis functions), one may wish 
to perform yet another study by a different all-electron method, in order 
to finally clarify the issue.

Our calculation setup can probably be further optimized for Fe systems 
by tuning the pseudopotential and extending the basis. 
So far, we managed to demonstrate that quite sensitive 
energetic and structural characteristics of Fe clusters can be reproduced
by a {\sc Siesta} method at a quite moderate computational cost.
Our next objective is the simulation of larger Fe particles for which, 
to our knowledge, structural relaxation has not yet been done. 

\section{Icosahedral vs. fcc clusters}
Clusters of icosahedral symmetry ($i$-) often come into discussion
for noble gases, simple metals and also transition metals.
In the present study, we simulated Fe$_{13}$ and Fe$_{55}$
$i$-particles as counterparts of fcc-structured clusters of corresponding size. 
Earlier results on $i$-Fe$_{13}$ have been published 
by Kortus \emph{et al.}\cite{APL80-4193}, and (among a number of isomers 
of Fe$_2$ -- Fe$_{17}$ clusters) -- 
by Di{\'e}guez \emph{et al.}\cite{PRB63-205407}. 
Duan and Zhang\cite{PhLA280-333} calculated electronic structure
of 13-at. and 55-at. icosahedral clusters of Fe, Co, and Ni by the
discrete variational method, and made a comparison with the results for
fcc-type and hexagonal isomers. For 13-at. clusters they optimized 
the cluster radius; the results for the 55-at. iron cluster correspond
to the interatomic distance fixed at that of bulk bcc iron.
Nevertheless some trends reported in Ref.~\onlinecite{PhLA280-333} hold in a
relaxed geometry and are discussed below. 

Our calculation for the $i$-Fe$_{13}$ in the GGA lead to a structure with
the total magnetic moment of 44 $\mu_{\mbox{\tiny B}}$, that is
identical to the all-electron result of Ref.~\onlinecite{APL80-4193}.
According to Duan and Zhang\cite{PhLA280-333}, the total moment is
46 $\mu_{\mbox{\tiny B}}$. 
We find a local magnetic moment 2.78 $\mu_{\mbox{\tiny B}}$ 
on the central atom, and 3.43 $\mu_{\mbox{\tiny B}}$ --
on its neighbors (separated by 2.43 {\AA} from the central one).
The bond length as relaxed in Ref.~\onlinecite{PhLA280-333} with the LDA
is slightly smaller -- 2.38 {\AA} (as could be generally expected
in the LDA, that is known to overestimate the binding).
Differently to Ref.~\onlinecite{APL80-4193} that reported an energy gap of
0.2 eV, the gap of only 0.05 eV was found in our case
(in the minority-spin subband, fully within a much larger gap of
0.6 eV in the majority-spin states).
The LDA calculation resulted in a smaller total moment 
of 36 $\mu_{\mbox{\tiny B}}$ in a somehow compressed
(2.30 {\AA} between central and peripheric atoms) cluster. 
The local magnetic moment of the central atom collapses to 
0.93 $\mu_{\mbox{\tiny B}}$. This makes a big difference to the
previously cited LDA result of Ref.~\onlinecite{PhLA280-333} in what
regards both the magnetization and the bond length. Such disagreement
is an indication that several metastable states of comparable
energy exist for the $i$--Fe$_{13}$ cluster, whereby the choice of
one or another happens due to minute differences in the calculation setup,
probably different starting conditions etc. A more systematic study
could be done as a sequence of fixed-moment calculations, as
discussed above for the Fe$_5$ cluster.
It is noteworthy that the recent study by the same {\sc Siesta} method
with only minutely different calculation setup \cite{PRB63-205407}
indicated the icosahedral cluster with not just reduced
but actually inverted magnetization
of the central atom (and total moment 34 $\mu_{\mbox{\tiny B}}$)
as the ground-state structure among
several competing isomers (all treated in the LDA only).
Such scattering of calculation results could be
a good indication that collinear ordering of local magnetic moments
becomes unstable. We tried to allow for non-collinear spin structure
at least for the Fe$_{13}$ cluster, but failed so far to arrive
at a reliably converged magnetic configuration within 
a reasonable calculation time.

The tendency for the reduction (or inversion) of the central magnetic moment 
in the icosahedral environment becomes more pronounced in a larger Fe$_{55}$ 
cluster. The magnetic moment of the central atom gets inverted 
($-$0.19 $\mu_{\mbox{\tiny B}}$) even in the GGA calculation, that otherwise 
favours ferromagnetic structure as we have seen above. Note that
Duan and Zheng\cite{PhLA280-333} also obtained an inverted spin moment 
on the central atom in their (unrelaxed) $i$-Fe$_{55}$ cluster, 
as calculated with the LDA.
The magnetic moments in the inner icosahedral shell are in our case 
suppressed to $\sim$0.34 $\mu_{\mbox{\tiny B}}$, and the moments 
in the outer shell reach merely 1.97 $\mu_{\mbox{\tiny B}}$ 
(for thirty 8-coordinated mid-edge atoms)
to 2.25 $\mu_{\mbox{\tiny B}}$ (twelve 6-coordinated vertex atoms).
The qualitative trend of how magnetic moments
change over icosahedral shells agrees with the results of 
Ref.~\onlinecite{PhLA280-333}. The difference in absolute numbers can be
related to a substantial compression we find in the relaxed cluster.
The radius of the inner shell shrinks to 2.4 {\AA}, i.e. by about 5\%
if compared to the equilibrium interatomic spacing in crystal.
In the case of fcc structure, such decrease in volume would change the 
ferromagnetic high-spine state into one of competing antiferromagnetic
arrangements as, e.g., Ref.~\onlinecite{PRB60-3839} shows,
or probably in a more complicated non-collinear structure,
with moderate local moments.
The relaxed radii of next spheres are about 4.2 {\AA} (30 atoms) and
4.9 {\AA} (12 atoms). 
Duan and Zheng\cite{PhLA280-333} did not take inner compression
in their $i$-cluster into account, that explains larger values of
magnetic moments they obtained.

One can conclude therefore that $i$-clusters do actually demand for additional
studies, where all existing ambiguities (ferromagnetic vs. ferrimagnetic 
ordering, GGA vs. LDA) will be analyzed on a more systematic basis,
like e.g. running a sequence of fixed-spin-moment calculations. In principle 
one could expect the presence of several structural solutions for some 
total moment numbers.

\section{Larger bcc and fcc-structured clusters}
\begin{figure}[p]
\centerline{\epsfig{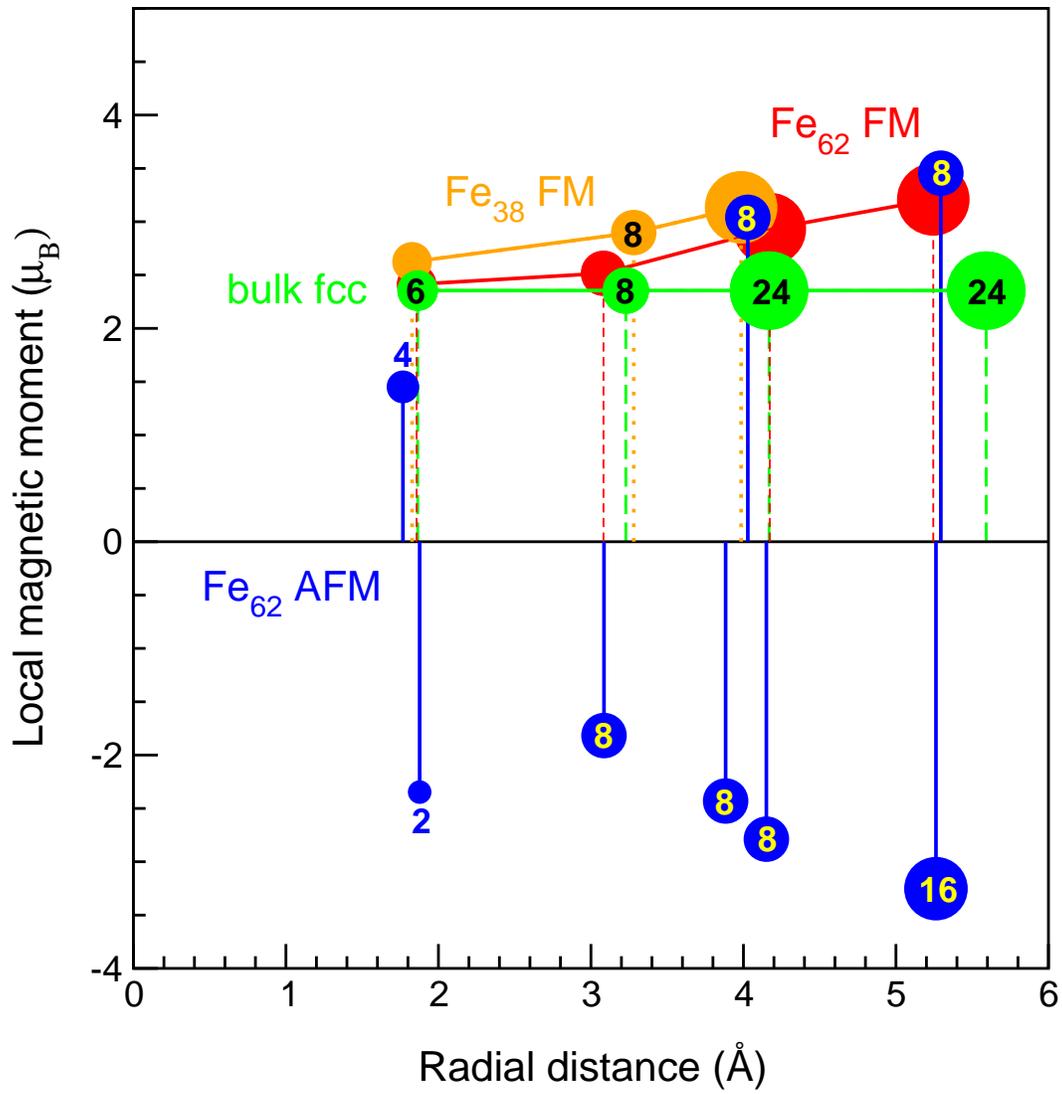}}
\bigskip
\caption{Distribution of local magnetic moments over shells
of neighbors in fcc-related nanoparticles Fe$_{38}$ and Fe$_{62}$,
centered around an octahedral interstitial. For the latter
particle, results corresponding to FM and AFM ordering
are shown. The size of the circle indicates the number of
neighbors a in corresponding shell. The data for
the fcc lattice are shown for comparison.}
\label{fig:uncentered}
\end{figure}

\begin{figure}[p]
\centerline{\epsfig{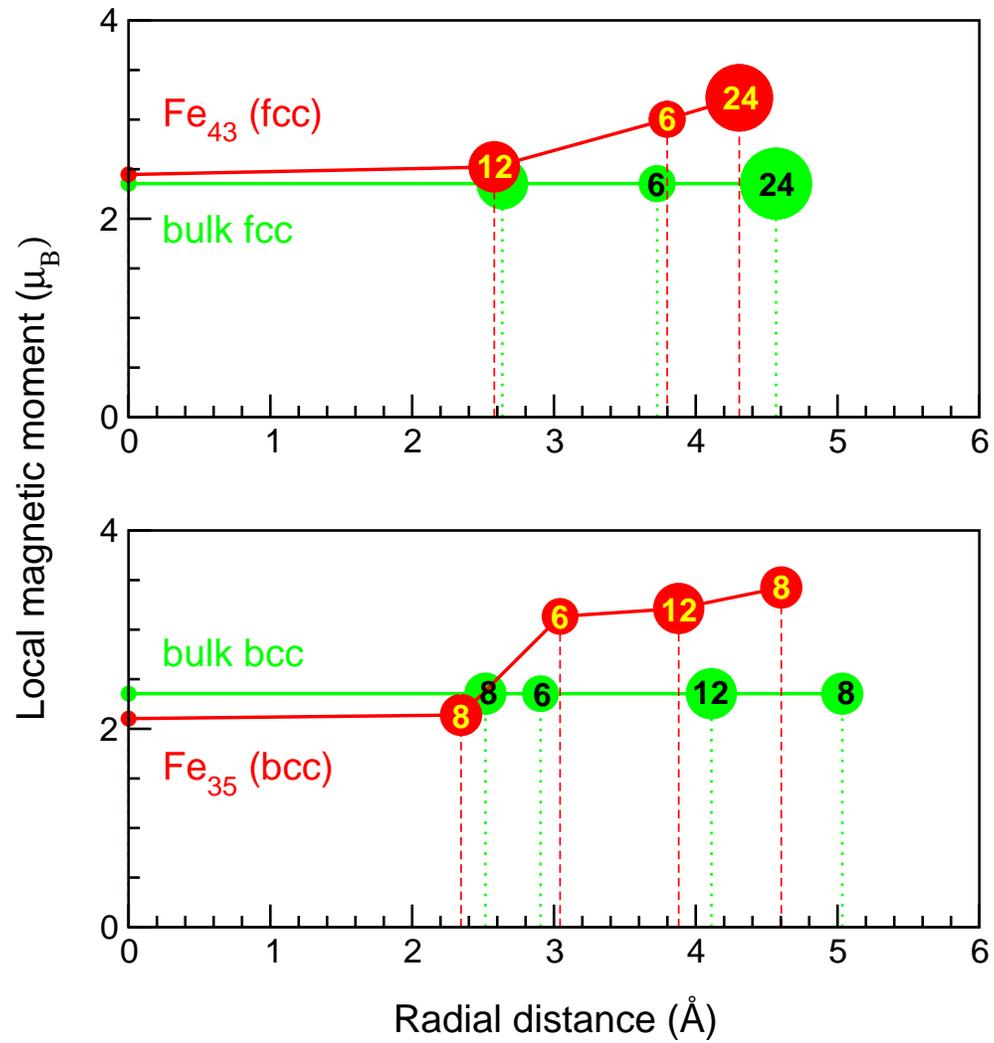}}
\bigskip
\caption{Same as in Fig.~\protect\ref{fig:uncentered},
for atom-centered Fe$_{35}$ (bcc) and Fe$_{43}$ (fcc)
particles. The data for corresponding perfect lattices
are shown for comparison.}
\label{fig:centered}
\end{figure}

The study of morphology and magnetic ordering of small clusters
is a delicate matter, sensitive to the calculation details and
prone to computational instabilities.
Corresponding results are accessible from experiment rather indirectly.
In larger metal nanoparticles (from $\sim$10$^2$
atoms on), ordered structures are often detected by electron microscopy
studies, so that substantial deviations from crystalline behavior remain
constrained to a (more or less thick) surface layer.
The aim of our further study was to simulate structure
relaxation and radial distribution of magnetic moments
inside Fe particles with several tens of atoms. Calculations
for bcc-related Fe$_{15}$ and Fe$_{35}$ clusters, albeit without
structure relaxation, have been reported by 
Fujima and Yamaguchi \cite{MSE217-295} (in a row of Ni and Cr
clusters of comparable size). Therefore we address in the following
primarily the fcc-related clusters, including Fe$_{35}$
for comparison. We considered particles having either a central
atom, or centered around an octahedral interstitial, and allowed unconstrained
structure relaxation after introducing small off-center displacements.
The relaxed structure essentially preserved the cubic point symmetry,
with the exception of the AFM (of CuAu-type) Fe$_{62}$ particle
that developed a slight tetragonal distortion.
The (relaxed) radii of atomic shells along with corresponding
magnetic moments are presented in Table~\ref{tab:Clusters}.
All these results are obtained in the GGA.
The values of local magnetic moments per atom are estimated
from the Mulliken population analysis. The values to be compared
to in perfect relaxed crystal, according to the {\sc Siesta}
calculation, are 2.43 $\mu_{\mbox{\tiny B}}$ (fcc)
and 2.35 $\mu_{\mbox{\tiny B}}$ (bcc). 
We show schematically the variation of magnetic moments
over relaxed spheres of neighbors in Fig.~\ref{fig:uncentered}
(for particles formed around an octahedral interstitial)
and Fig.~\ref{fig:centered} (for particles with a central atom).

%
%
\begin{table}[b]
\caption{Relaxed distances from center $a$ and magnetic moments 
over shells of neighbors $M$ in bcc- and fcc-related Fe clusters.
Numbers of neighbors within each shell are given in parentheses.}
\begin{ruledtabular}
\begin{tabular}{cddccdd}
 & \multicolumn{1}{c}{$a$ ({\AA})} 
 & \multicolumn{1}{c}{$M$ ($\mu_{\mbox{\tiny B}}$)} &&
 & \multicolumn{1}{c}{$a$ ({\AA})} 
 & \multicolumn{1}{c}{$M$ ($\mu_{\mbox{\tiny B}}$)} \\
\multicolumn{3}{c}{\hrulefill~~~Fe$_{35}$ (bcc)~~~\hrulefill} && 
\multicolumn{3}{c}{\hrulefill~~~Fe$_{59}$ (bcc)~~~\hrulefill}\\
 (1)& 0.0    &    2.10 &&  (1)& 0.0    &  2.85 \\
 (8)& 2.345  &    2.14 &&  (8)& 2.525  &  2.64 \\
 (6)& 3.043  &    3.13 &&  (6)& 3.202  &  2.45 \\
(12)& 3.880  &    3.21 && (12)& 4.026  &  2.79 \\
 (8)& 4.603  &    3.43 && (24)& 4.678  &  3.25 \\
\multicolumn{3}{c}{\hrulefill~~~Fe$_{38}$ (fcc)~~~\hrulefill} && 
                           (8)& 4.821  &  3.13 \\
 (6)& 1.827  &    2.62 &\multicolumn{1}{c}{}&  
\multicolumn{3}{c}{\hrulefill~~~Fe$_{62}$ (fcc)~~~\hrulefill}\\
 (8)& 3.281  &    2.90 &&  (6)& 1.857  &  2.41 \\
(24)& 3.985  &    3.14 &&  (8)& 3.082  &  2.52 \\
\multicolumn{3}{c}{\hrulefill~~~Fe$_{43}$ (fcc)~~~\hrulefill} && 
                          (24)& 4.173  &  2.93 \\
 (1)& 0.0    &    2.45 && (24)& 5.245  &  3.21 \\
(12)& 2.579  &    2.52 &\multicolumn{1}{c}{}& 
\multicolumn{3}{c}{\hrulefill~~~Fe$_{62}$ (fcc), AFM~~~\hrulefill}\\
 (6)& 3.798  &    3.00 &&  (4)& 1.767  &  1.45 \\
(24)& 4.307  &    3.22 &&  (2)& 1.877  & -2.35 \\
\multicolumn{3}{c}{\hrulefill~~~Fe$_{55}$ (fcc)~~~\hrulefill} && 
                           (8)& 3.086  & -1.82 \\
 (1)& 0.0    &    2.29 &&  (8)& 3.883  & -2.43 \\
(12)& 2.507  &    2.26 &&  (8)& 4.029  &  3.04 \\
 (6)& 3.361  &    2.73 &&  (8)& 4.150  & -2.79 \\
(24)& 4.124  &    2.85 && (16)& 5.262  & -3.25 \\
(12)& 4.797  &    3.17 &&  (8)& 5.294  &  3.46 \\
\end{tabular}
\end{ruledtabular}
\label{tab:Clusters}
\end{table}

The comparison of unrelaxed bulk-like bcc geometry with
the final relaxed structure helps to attribute the drop
of the magnetic moment in the first (8-at.) coordination shell
around the central atom in the (bcc) Fe$_{35}$ cluster,
found by Fujima and Yamaguchi \cite{MSE217-295},
to the structure effect (and not, say, to the differences
in the calculation scheme or in the definition of local magnetic moments).
If the neighbors' positions are fixed as in the bulk, 
the local magnetic moments are 2.31 (on the central atom); 2.21; 2.98; 
3.23 and 3.45 (on the surface) $\mu_{\mbox{\tiny B}}$, i.e.
the drop in the magnetic moment of the second shell is reproduced. 
However, with full relaxation taken into account in the Fe$_{35}$ cluster,
we found the magnetic moment to grow steadily towards the surface
(see Fig.~\ref{fig:centered}, Table \ref{tab:Clusters}).
This effect seems to be intrinsic to the bcc morphology since we again 
find it in a larger Fe$_{59}$ cluster:
the inward relaxation of the atoms in the first coordination shell
gradually increases their local magnetic moments.
In this process, the radii of 8-at. and 6-at. coordination spheres,
whose relation in the bulk is nearly 0.87,
become pronouncedly separated in clusters, reducing the above ratio
to 0.77 in Fe$_{35}$ and 0.79 in Fe$_{59}$ (see Fig.~\ref{fig:centered}, 
bottom panel).

Consistently with the results of Ref.~\onlinecite{MSE217-295}, 
the magnetic moments grow in the outer shells and exceed
3 $\mu_{\mbox{\tiny B}}$ on the surface.
The remaining quantitative differences between our results
for the Fe$_{35}$ cluster and those
by Fujima and Yamagichi~\cite{MSE217-295}
can be due to different exchange-correlation
potential ($X_{\alpha}$ used in Ref.~\onlinecite{MSE217-295}). 
The relaxation (neglected in Ref.~\onlinecite{MSE217-295})
is outwards for the second shell (6 atoms) and inwards
in all others. We found (this applies to the fcc clusters as well)
the inward relaxation to be the largest on the surface,
where the magnetic moment is at most enhanced. 
This behaviour is in consistence with well-known trends at
the surface of bulk Fe. The enhancement of \emph{local}
magnetic moments (i.e., inside muffin-tin spheres) in the slab
full-potential calculation was found by Freeman and Fu \cite{JAP61-3356}
to be from 2.15 $\mu_{\mbox{\tiny B}}$ (bulk) to 2.65 $\mu_{\mbox{\tiny B}}$
at the (110) surface and 2.98 $\mu_{\mbox{\tiny B}}$ at the (100) surface.
The surface relaxation (see, e.g., a recent first-principle calculation
by Spencer \emph{et al.}\cite{SS513-389}, that also reviews 
experimental results for different Fe surfaces) is always inwards 
for the upper layer.

In the fcc-related clusters, the internal structure is more
densely packed, and a pronounced outward relaxation
occurs for the second shell (6 at.) of the atom-centered
Fe$_{43}$ cluster. Comparing this to the result for the
(interstitial-centered) Fe$_{62}$ cluster, one sees moreover a general tendency
of developing a (quite small) outward relaxation in the subsurface
shell, whereas the outer shell is always strongly contracted.
Towards further inner shells, the relaxation is rapidly stabilized,
and the atomic spacing approaches that in the bulk.
As in the case of the bcc cluster, magnetic moments are largely
enhanced on the surface (and immediately below it), but rapidly decrease and
get stabilized in the deeper shells, without showing any fluctuations.
Such behaviour agrees qualitatively with the results of
Duan and Zheng\cite{PhLA280-333} on the fcc Fe$_{55}$ cluster,
done however with fixed interatomic distances (equal to those
in the bcc bulk). We do not find any clear reversal of
magnetization on the central atom of the fcc Fe$_{55}$ cluster,
in contrary to what was reported in Ref.~\onlinecite{PhLA280-333}.
One cannot exclude the possibility for such a configuration
to emerge in one of metastable states, in a calculation
departing from a specially prepared initial magnetic configuration.
However, whereas AFM and FM types of ordering are known to be competitive
in fcc Fe, on the high-spin side (large volumes) the FM arrangement
is definitely more favourable, at least in the bulk\cite{PRB60-3839}.
Moreover, experimental estimates of mean magnetic moments for the
Fe$_{55}$ cluster (3.1 $\mu_{\mbox{\tiny B}}$, according to
Ref.~\onlinecite{PRL71-4067}) are in better agreement with our 
result of 2.77 $\mu_{\mbox{\tiny B}}$ than with 2.58 $\mu_{\mbox{\tiny B}}$
after Duan and Zheng\cite{PhLA280-333}.

An additional study of the plausibility of AFM ordering was undertaken
for the Fe$_{62}$ cluster. The layer-by-layer (i.e., CuAu-type)
AFM organization lifts some degeneracies
in the radial distribution of atoms (see Fig.~\ref{fig:uncentered}),
but otherwise is consistent
with the above observations (inward surface relaxation;
enhancement of magnetic moments towards the surface).
The total energy per atom is by 0.19 eV higher in the AFM configuration
than in the corresponding FM case, that effectively
rules out this particular (admittedly arbitrary) magnetic configuration 
as a competitive one. However, the fact that the magnitudes of magnetic 
moments over shells are almost identical in AFM and FM cases implies
that the individual spins are localized enough to survive
in different magnetic orientations, and could probably prove
a gain in the magnetic energy in some of them.

It could be instructive to analyze how the enhanced magnetization
at the surface decomposes into contributions from different basis
orbitals, and how the electron density gets distributed between the inner 
part and the surface of clusters in the course of relaxation.
Fortunately one can establish very clear trends, common for
fcc and bcc clusters of different size. 
Magnetic polarization of the $4s$ states is in all cases well below
0.1 $\mu_{\mbox{\tiny B}}$ and antiparallel to the magnetic moment
in the $3d$ shell. It gradually disappears 
(to about 0.01 $\mu_{\mbox{\tiny B}}$) towards the surface.
This is consistent with the magnetic moment decomposition
for the Fe$_{55}$ cluster by Duan and Zheng\cite{PhLA280-333}. 
In disagreement with the latter, we find that the polarization 
in the $4p$ shells, being parallel to that of $3d$, steadily increases 
from the core region outwards and contributes as much
as 7\% of local magnetic moments at the surface.
The major, $3d$-related, part of the local moment grows 
due to simultaneous increase of majority-spin and decrease
of minority-spin occupations. However, the majority-spin $3d$ subband
never gets saturated (in fact, its occupation does not exceed 4.8 electrons).
The charge transfer always happen from the core region to the surface;
the nominal valence charge per Fe atom is 7.4--7.6 in the cluster core
and slightly beyond 8 at the surface. As definition-dependent
as these qualitative estimates might be, they do not let to oversee
a general qualitative trend. Duan and Zheng\cite{PhLA280-333}
found fluctuations of charge from one atom shell to another,
but on the average no clear distinction between core and
surface atoms -- hence the difference we talk about must come from relaxation,
taken into account in our case. It is understandable that the
overall compression in the cluster due to its ``surface tension''
shifts upwards the electron states in the inner region, that results 
in the outward charge flow.
Surface atoms experience stronger relaxation but they have lower
coordination. The atoms adjust in the cluster so as to smooth
the radial charge distribution; simultaneously the magnetization
profile tends to acquire certain regularity. The enhancement
of the $4p$ contribution to the magnetic polarization on the surface
is related with the abovementioned redistribution of charge,
because extra electrons in the surface layer can be more easily
accommodated by majority-spin states of the $p$ symmetry.

According to the experimental evidence \cite{PRL71-4067,Sci265-1682},
average magnetic moments per atom start from nearly 
3 $\mu_{\mbox{\tiny B}}$ in small clusters and gradually
decrease to the bulk value of 2.2 $\mu_{\mbox{\tiny B}}$ in
400--500 atom particles. This implies ferromagnetic ordering
and bcc-related structure in large clusters. The fluctuations
of magnetization prior to this asymptotic value being achieved
are not yet systematically explained. Billas \emph{et al.}
emphasize \cite{PRL71-4067} that the mean magnetic moment 
in the Fe$_{55}$ atom is anomalously large
(nearly 3.1 $\mu_{\mbox{\tiny B}}$), and bring the
icosahedral structure into discussion. However, we have seen that
the $i$-Fe$_{55}$ cluster has in fact a moderate mean magnetic moment
of 2.2 $\mu_{\mbox{\tiny B}}$ (or slightly larger, but anyway
well below that of the fcc cluster, in Ref.~\onlinecite{PhLA280-333}). 
In order to describe experimental variations of mean magnetic moments
in simple terms, Billas \emph{et al.} proposed \cite{Sci265-1682}
a model of Fe magnetic moments decreasing from the surface
into the interior of the cluster, and getting inverted in the fourth
shell from the surface. While essentially confirming this model in
what concerns the asymptotic end values of magnetic moments
(on the surface and in the bcc interior),
our simulation does not yet produce any strong evidence
for the fluctuations in between, when starting from
ferromagnetic test configuration.
However, a more detailed analysis (e.g., within a fixed spin
moment scheme) could help to single out other competing
magnetic structures.

Addressing the issue of structural order vs. disorder,
it is worth noting that
Soler \emph{et al.} \cite{PRB61-5771} studied the morphology
of small (38 to 75 atoms) ``ordered'' and ``amorphous''
gold nanoparticles, using the same calculation scheme as here 
in order to refine trial geometries, provided by energy minimization 
in an empirical potential. In all cases ``amorphous'' particles
were found to be more stable, and the subsequent analysis 
shows the reason for this to be due to high elastic contribution
to the total energy of a particle, relaxed in the course
of an amorphous-like rearrangement of atoms. In our case,
inner shells of Fe particles are also contracted, as compared
to the bulk crystal. It would be interesting to probe
the effects of amorphization, and their interplay with
magnetic characteristics, in an \emph{ab initio} simulation
once realistic models for ground-state Fe arrangements
become available.

Summarizing, our results favors a conclusion that the relaxation and
magnetic properties of small Fe nanoparticles have certain
common features, relatively independent on morphology, magnetic
ordering and size. Namely, the structure relaxation is practically
confined within 2--3 outer shells, the surface layer relaxes
strongly inward, and the magnetic moments on the surface are
enhanced to beyond 3 $\mu_{\mbox{\tiny B}}$. 
The overall magnetic properties of larger nanoparticles must be then
primarily governed by the proportion between surface-layers atoms
and their deep bulk-like counterparts. 

\section*{Acknowledgments}
The work was supported by the German Research Society (SFB 445)
and by Spain's Fundaci\'on Ram\'on Areces and
MCyT (BFM2000-1312).
The authors are grateful to L.~C.~Balb{\'a}s and P.~Ordej\'on
for useful discussions. JMS thanks J.~K\"ubler and
R.~Martin for their help in the development of the non-collinear
magnetization option in {\sc Siesta}. AVP thanks
J.~Kortus for useful discussions on the subject of
small Fe clusters.


\begin{thebibliography}{46}
\expandafter\ifx\csname natexlab\endcsname\relax\def\natexlab#1{#1}\fi
\expandafter\ifx\csname bibnamefont\endcsname\relax
  \def\bibnamefont#1{#1}\fi
\expandafter\ifx\csname bibfnamefont\endcsname\relax
  \def\bibfnamefont#1{#1}\fi
\expandafter\ifx\csname citenamefont\endcsname\relax
  \def\citenamefont#1{#1}\fi
\expandafter\ifx\csname url\endcsname\relax
  \def\url#1{\texttt{#1}}\fi
\expandafter\ifx\csname urlprefix\endcsname\relax\def\urlprefix{URL }\fi
\providecommand{\bibinfo}[2]{#2}
\providecommand{\eprint}[2][]{\url{#2}}

\bibitem[{\citenamefont{Billas et~al.}(1993)\citenamefont{Billas, Becker,
  Ch{\^a}telain, and {de}~Heer}}]{PRL71-4067}
\bibinfo{author}{\bibfnamefont{I.~M.~L.} \bibnamefont{Billas}},
  \bibinfo{author}{\bibfnamefont{J.~A.} \bibnamefont{Becker}},
  \bibinfo{author}{\bibfnamefont{A.}~\bibnamefont{Ch{\^a}telain}},
  \bibnamefont{and} \bibinfo{author}{\bibfnamefont{W.~A.}
  \bibnamefont{{de}~Heer}}, \bibinfo{journal}{Phys.~Rev.~Lett.}
  \textbf{\bibinfo{volume}{71}}, \bibinfo{pages}{4067} (\bibinfo{year}{1993}).

\bibitem[{\citenamefont{Billas et~al.}(1994)\citenamefont{Billas,
  Ch{\^{a}}telain, and {de}~Heer}}]{Sci265-1682}
\bibinfo{author}{\bibfnamefont{I.~M.~L.} \bibnamefont{Billas}},
  \bibinfo{author}{\bibfnamefont{A.}~\bibnamefont{Ch{\^{a}}telain}},
  \bibnamefont{and} \bibinfo{author}{\bibfnamefont{W.~A.}
  \bibnamefont{{de}~Heer}}, \bibinfo{journal}{Science}
  \textbf{\bibinfo{volume}{265}}, \bibinfo{pages}{1682} (\bibinfo{year}{1994}).

\bibitem[{\citenamefont{Gerion et~al.}(2000)\citenamefont{Gerion, Hirt, Billas,
  Ch{\^{a}}telain, and {de Heer}}}]{PRB62-7491}
\bibinfo{author}{\bibfnamefont{D.}~\bibnamefont{Gerion}},
  \bibinfo{author}{\bibfnamefont{A.}~\bibnamefont{Hirt}},
  \bibinfo{author}{\bibfnamefont{I.~M.~L.} \bibnamefont{Billas}},
  \bibinfo{author}{\bibfnamefont{A.}~\bibnamefont{Ch{\^{a}}telain}},
  \bibnamefont{and} \bibinfo{author}{\bibfnamefont{W.~A.} \bibnamefont{{de
  Heer}}}, \bibinfo{journal}{Phys.~Rev.~B} \textbf{\bibinfo{volume}{62}},
  \bibinfo{pages}{7491} (\bibinfo{year}{2000}).

\bibitem[{\citenamefont{Christensen and Cohen}(1993)}]{PRB47-13643}
\bibinfo{author}{\bibfnamefont{O.~B.} \bibnamefont{Christensen}}
  \bibnamefont{and} \bibinfo{author}{\bibfnamefont{M.~L.} \bibnamefont{Cohen}},
  \bibinfo{journal}{Phys.~Rev.~B} \textbf{\bibinfo{volume}{47}},
  \bibinfo{pages}{13643} (\bibinfo{year}{1993}).

\bibitem[{\citenamefont{Besley et~al.}(1995)\citenamefont{Besley, Johnston,
  Stace, and Uppenbrink}}]{JMC341-75}
\bibinfo{author}{\bibfnamefont{N.~A.} \bibnamefont{Besley}},
  \bibinfo{author}{\bibfnamefont{R.~L.} \bibnamefont{Johnston}},
  \bibinfo{author}{\bibfnamefont{A.~J.} \bibnamefont{Stace}}, \bibnamefont{and}
  \bibinfo{author}{\bibfnamefont{J.}~\bibnamefont{Uppenbrink}},
  \bibinfo{journal}{Journal of Molecular Structure (Theochem)}
  \textbf{\bibinfo{volume}{341}}, \bibinfo{pages}{75} (\bibinfo{year}{1995}).

\bibitem[{\citenamefont{Pastor et~al.}(1989)\citenamefont{Pastor,
  Dorantes-D{\'a}vila, and Bennemann}}]{PRB40-7642}
\bibinfo{author}{\bibfnamefont{G.~M.} \bibnamefont{Pastor}},
  \bibinfo{author}{\bibfnamefont{J.}~\bibnamefont{Dorantes-D{\'a}vila}},
  \bibnamefont{and} \bibinfo{author}{\bibfnamefont{K.~H.}
  \bibnamefont{Bennemann}}, \bibinfo{journal}{Phys.~Rev.~B}
  \textbf{\bibinfo{volume}{40}}, \bibinfo{pages}{7642} (\bibinfo{year}{1989}).

\bibitem[{\citenamefont{Pastor et~al.}(1996)\citenamefont{Pastor, Hirsch, and
  M{\"u}hlschlegel}}]{PRB53-10382}
\bibinfo{author}{\bibfnamefont{G.~M.} \bibnamefont{Pastor}},
  \bibinfo{author}{\bibfnamefont{R.}~\bibnamefont{Hirsch}}, \bibnamefont{and}
  \bibinfo{author}{\bibfnamefont{B.}~\bibnamefont{M{\"u}hlschlegel}},
  \bibinfo{journal}{Phys.~Rev.~B} \textbf{\bibinfo{volume}{53}},
  \bibinfo{pages}{10382} (\bibinfo{year}{1996}).

\bibitem[{\citenamefont{Andriotis
  et~al.}(1996{\natexlab{a}})\citenamefont{Andriotis, Lathiotakis, and
  Menon}}]{ChPL260-15}
\bibinfo{author}{\bibfnamefont{A.~N.} \bibnamefont{Andriotis}},
  \bibinfo{author}{\bibfnamefont{N.}~\bibnamefont{Lathiotakis}},
  \bibnamefont{and} \bibinfo{author}{\bibfnamefont{M.}~\bibnamefont{Menon}},
  \bibinfo{journal}{Chem.~Phys.~Lett.} \textbf{\bibinfo{volume}{260}},
  \bibinfo{pages}{15} (\bibinfo{year}{1996}{\natexlab{a}}).

\bibitem[{\citenamefont{Andriotis
  et~al.}(1996{\natexlab{b}})\citenamefont{Andriotis, Lathiotakis, and
  Menon}}]{EPL36-37}
\bibinfo{author}{\bibfnamefont{A.~N.} \bibnamefont{Andriotis}},
  \bibinfo{author}{\bibfnamefont{N.~N.} \bibnamefont{Lathiotakis}},
  \bibnamefont{and} \bibinfo{author}{\bibfnamefont{M.}~\bibnamefont{Menon}},
  \bibinfo{journal}{Europhys.~Lett.} \textbf{\bibinfo{volume}{36}},
  \bibinfo{pages}{37} (\bibinfo{year}{1996}{\natexlab{b}}).

\bibitem[{\citenamefont{Andriotis and Menon}(1998)}]{PRB57-10069}
\bibinfo{author}{\bibfnamefont{A.~N.} \bibnamefont{Andriotis}}
  \bibnamefont{and} \bibinfo{author}{\bibfnamefont{M.}~\bibnamefont{Menon}},
  \bibinfo{journal}{Phys.~Rev.~B} \textbf{\bibinfo{volume}{57}},
  \bibinfo{pages}{10069} (\bibinfo{year}{1998}).

\bibitem[{\citenamefont{Bouarab et~al.}(1996)\citenamefont{Bouarab, Vega,
  Alonso, and I{\~{n}}iguez}}]{PRB54-3003}
\bibinfo{author}{\bibfnamefont{S.}~\bibnamefont{Bouarab}},
  \bibinfo{author}{\bibfnamefont{A.}~\bibnamefont{Vega}},
  \bibinfo{author}{\bibfnamefont{J.~A.} \bibnamefont{Alonso}},
  \bibnamefont{and} \bibinfo{author}{\bibfnamefont{M.~P.}
  \bibnamefont{I{\~{n}}iguez}}, \bibinfo{journal}{Phys.~Rev.~B}
  \textbf{\bibinfo{volume}{54}}, \bibinfo{pages}{3003} (\bibinfo{year}{1996}).

\bibitem[{\citenamefont{Guevara et~al.}(1997)\citenamefont{Guevara, Parisi,
  Llois, and Weissmann}}]{PRB55-13283}
\bibinfo{author}{\bibfnamefont{J.}~\bibnamefont{Guevara}},
  \bibinfo{author}{\bibfnamefont{F.}~\bibnamefont{Parisi}},
  \bibinfo{author}{\bibfnamefont{A.~M.} \bibnamefont{Llois}}, \bibnamefont{and}
  \bibinfo{author}{\bibfnamefont{M.}~\bibnamefont{Weissmann}},
  \bibinfo{journal}{Phys.~Rev.~B} \textbf{\bibinfo{volume}{55}},
  \bibinfo{pages}{13283} (\bibinfo{year}{1997}).

\bibitem[{\citenamefont{Kohn}(1999)}]{RMP71-1253}
\bibinfo{author}{\bibfnamefont{W.}~\bibnamefont{Kohn}},
  \bibinfo{journal}{Rev.~Mod.~Phys.} \textbf{\bibinfo{volume}{71}},
  \bibinfo{pages}{1253} (\bibinfo{year}{1999}).

\bibitem[{\citenamefont{Chen et~al.}(1991)\citenamefont{Chen, Wang, Jackson,
  and Pederson}}]{PRB44-6558}
\bibinfo{author}{\bibfnamefont{J.~L.} \bibnamefont{Chen}},
  \bibinfo{author}{\bibfnamefont{C.~S.} \bibnamefont{Wang}},
  \bibinfo{author}{\bibfnamefont{K.~A.} \bibnamefont{Jackson}},
  \bibnamefont{and} \bibinfo{author}{\bibfnamefont{M.~R.}
  \bibnamefont{Pederson}}, \bibinfo{journal}{Phys.~Rev.~B}
  \textbf{\bibinfo{volume}{44}}, \bibinfo{pages}{6558} (\bibinfo{year}{1991}).

\bibitem[{\citenamefont{Castro and Salahub}(1994)}]{PRB49-11842}
\bibinfo{author}{\bibfnamefont{M.}~\bibnamefont{Castro}} \bibnamefont{and}
  \bibinfo{author}{\bibfnamefont{D.~R.} \bibnamefont{Salahub}},
  \bibinfo{journal}{Phys.~Rev.~B} \textbf{\bibinfo{volume}{49}},
  \bibinfo{pages}{11842} (\bibinfo{year}{1994}).

\bibitem[{\citenamefont{Castro et~al.}(1997)\citenamefont{Castro, Jamorski, and
  Salahub}}]{ChPL271-133}
\bibinfo{author}{\bibfnamefont{M.}~\bibnamefont{Castro}},
  \bibinfo{author}{\bibfnamefont{C.}~\bibnamefont{Jamorski}}, \bibnamefont{and}
  \bibinfo{author}{\bibfnamefont{D.~R.} \bibnamefont{Salahub}},
  \bibinfo{journal}{Chem.~Phys.~Lett.} \textbf{\bibinfo{volume}{271}},
  \bibinfo{pages}{133} (\bibinfo{year}{1997}).

\bibitem[{\citenamefont{Kortus et~al.}(2002)\citenamefont{Kortus, Baruah,
  Pederson, Ashman, and Khanna}}]{APL80-4193}
\bibinfo{author}{\bibfnamefont{J.}~\bibnamefont{Kortus}},
  \bibinfo{author}{\bibfnamefont{T.}~\bibnamefont{Baruah}},
  \bibinfo{author}{\bibfnamefont{M.~R.} \bibnamefont{Pederson}},
  \bibinfo{author}{\bibfnamefont{C.}~\bibnamefont{Ashman}}, \bibnamefont{and}
  \bibinfo{author}{\bibfnamefont{S.~N.} \bibnamefont{Khanna}},
  \bibinfo{journal}{Appl.~Phys.~Lett.} \textbf{\bibinfo{volume}{80}},
  \bibinfo{pages}{4193} (\bibinfo{year}{2002}).

\bibitem[{\citenamefont{Ballone and Jones}(1995)}]{ChPL233-632}
\bibinfo{author}{\bibfnamefont{P.}~\bibnamefont{Ballone}} \bibnamefont{and}
  \bibinfo{author}{\bibfnamefont{R.~O.} \bibnamefont{Jones}},
  \bibinfo{journal}{Chem.~Phys.~Lett.} \textbf{\bibinfo{volume}{233}},
  \bibinfo{pages}{632} (\bibinfo{year}{1995}).

\bibitem[{\citenamefont{Oda et~al.}(1998)\citenamefont{Oda, Pasquarello, and
  Car}}]{PRL80-3622}
\bibinfo{author}{\bibfnamefont{T.}~\bibnamefont{Oda}},
  \bibinfo{author}{\bibfnamefont{A.}~\bibnamefont{Pasquarello}},
  \bibnamefont{and} \bibinfo{author}{\bibfnamefont{R.}~\bibnamefont{Car}},
  \bibinfo{journal}{Phys.~Rev.~Lett.} \textbf{\bibinfo{volume}{80}},
  \bibinfo{pages}{3622} (\bibinfo{year}{1998}).

\bibitem[{\citenamefont{Hobbs et~al.}(2000)\citenamefont{Hobbs, Kresse, and
  Hafner}}]{PRB62-11556}
\bibinfo{author}{\bibfnamefont{D.}~\bibnamefont{Hobbs}},
  \bibinfo{author}{\bibfnamefont{G.}~\bibnamefont{Kresse}}, \bibnamefont{and}
  \bibinfo{author}{\bibfnamefont{J.}~\bibnamefont{Hafner}},
  \bibinfo{journal}{Phys.~Rev.~B} \textbf{\bibinfo{volume}{62}},
  \bibinfo{pages}{11556} (\bibinfo{year}{2000}).

\bibitem[{\citenamefont{Alonso}(2000)}]{ChemRev100-637}
\bibinfo{author}{\bibfnamefont{J.~A.} \bibnamefont{Alonso}},
  \bibinfo{journal}{Chem. Rev.} \textbf{\bibinfo{volume}{100}},
  \bibinfo{pages}{637} (\bibinfo{year}{2000}).

\bibitem[{\citenamefont{Fujima and Yamaguchi}(1996)}]{MSE217-295}
\bibinfo{author}{\bibfnamefont{N.}~\bibnamefont{Fujima}} \bibnamefont{and}
  \bibinfo{author}{\bibfnamefont{T.}~\bibnamefont{Yamaguchi}},
  \bibinfo{journal}{Materials Science and Engineering~A}
  \textbf{\bibinfo{volume}{217--218}}, \bibinfo{pages}{295}
  (\bibinfo{year}{1996}).

\bibitem[{\citenamefont{Duan and Zheng}(2001)}]{PhLA280-333}
\bibinfo{author}{\bibfnamefont{H.~M.} \bibnamefont{Duan}} \bibnamefont{and}
  \bibinfo{author}{\bibfnamefont{Q.~Q.} \bibnamefont{Zheng}},
  \bibinfo{journal}{Phys.~Lett.~A} \textbf{\bibinfo{volume}{280}},
  \bibinfo{pages}{333} (\bibinfo{year}{2001}).

\bibitem[{\citenamefont{S{\'a}nchez-Portal
  et~al.}(1997)\citenamefont{S{\'a}nchez-Portal, Ordej{\'o}n, Artacho, and
  Soler}}]{IJQC65-453}
\bibinfo{author}{\bibfnamefont{D.}~\bibnamefont{S{\'a}nchez-Portal}},
  \bibinfo{author}{\bibfnamefont{P.}~\bibnamefont{Ordej{\'o}n}},
  \bibinfo{author}{\bibfnamefont{E.}~\bibnamefont{Artacho}}, \bibnamefont{and}
  \bibinfo{author}{\bibfnamefont{J.~M.} \bibnamefont{Soler}},
  \bibinfo{journal}{Int. J. Quant. Chem.} \textbf{\bibinfo{volume}{65}},
  \bibinfo{pages}{453} (\bibinfo{year}{1997}).

\bibitem[{\citenamefont{Artacho et~al.}(1999)\citenamefont{Artacho,
  S{\'a}nchez-Portal, Ordej{\'o}n, Garc{\'{\i}}a, and Soler}}]{PSSB215-809}
\bibinfo{author}{\bibfnamefont{E.}~\bibnamefont{Artacho}},
  \bibinfo{author}{\bibfnamefont{D.}~\bibnamefont{S{\'a}nchez-Portal}},
  \bibinfo{author}{\bibfnamefont{P.}~\bibnamefont{Ordej{\'o}n}},
  \bibinfo{author}{\bibfnamefont{A.}~\bibnamefont{Garc{\'{\i}}a}},
  \bibnamefont{and} \bibinfo{author}{\bibfnamefont{J.~M.} \bibnamefont{Soler}},
  \bibinfo{journal}{Physica Status Solidi (b)} \textbf{\bibinfo{volume}{215}},
  \bibinfo{pages}{809} (\bibinfo{year}{1999}).

\bibitem[{\citenamefont{Soler et~al.}(2002)\citenamefont{Soler, Artacho, Gale,
  Garc{\'{\i}}a, Junquera, Ordej{\'o}n, and S{\'a}nchez-Portal}}]{JPCM14-2745}
\bibinfo{author}{\bibfnamefont{J.~M.} \bibnamefont{Soler}},
  \bibinfo{author}{\bibfnamefont{E.}~\bibnamefont{Artacho}},
  \bibinfo{author}{\bibfnamefont{J.~D.} \bibnamefont{Gale}},
  \bibinfo{author}{\bibfnamefont{A.}~\bibnamefont{Garc{\'{\i}}a}},
  \bibinfo{author}{\bibfnamefont{J.}~\bibnamefont{Junquera}},
  \bibinfo{author}{\bibfnamefont{P.}~\bibnamefont{Ordej{\'o}n}},
  \bibnamefont{and}
  \bibinfo{author}{\bibfnamefont{D.}~\bibnamefont{S{\'a}nchez-Portal}},
  \bibinfo{journal}{J.~Phys.:~Condens.~Matter} \textbf{\bibinfo{volume}{14}},
  \bibinfo{pages}{2745} (\bibinfo{year}{2002}).

\bibitem[{\citenamefont{Soler et~al.}(2000)\citenamefont{Soler, Beltr{\'a}n,
  Michaelian, Garz{\'o}n, Ordej{\'o}n, {S{\'a}nchez-Portal}, and
  Artacho}}]{PRB61-5771}
\bibinfo{author}{\bibfnamefont{J.~M.} \bibnamefont{Soler}},
  \bibinfo{author}{\bibfnamefont{M.~R.} \bibnamefont{Beltr{\'a}n}},
  \bibinfo{author}{\bibfnamefont{K.}~\bibnamefont{Michaelian}},
  \bibinfo{author}{\bibfnamefont{I.~L.} \bibnamefont{Garz{\'o}n}},
  \bibinfo{author}{\bibfnamefont{P.}~\bibnamefont{Ordej{\'o}n}},
  \bibinfo{author}{\bibfnamefont{D.}~\bibnamefont{{S{\'a}nchez-Portal}}},
  \bibnamefont{and} \bibinfo{author}{\bibfnamefont{E.}~\bibnamefont{Artacho}},
  \bibinfo{journal}{Phys.~Rev.~B} \textbf{\bibinfo{volume}{61}},
  \bibinfo{pages}{5771} (\bibinfo{year}{2000}).

\bibitem[{\citenamefont{Izquierdo et~al.}(2000)\citenamefont{Izquierdo, Vega,
  Balb{\'a}s, S{\'a}nchez-Portal, Junquera, Artacho, Soler, and
  Ordej{\'o}n}}]{PRB61-13639}
\bibinfo{author}{\bibfnamefont{J.}~\bibnamefont{Izquierdo}},
  \bibinfo{author}{\bibfnamefont{A.}~\bibnamefont{Vega}},
  \bibinfo{author}{\bibfnamefont{L.~C.} \bibnamefont{Balb{\'a}s}},
  \bibinfo{author}{\bibfnamefont{D.}~\bibnamefont{S{\'a}nchez-Portal}},
  \bibinfo{author}{\bibfnamefont{J.}~\bibnamefont{Junquera}},
  \bibinfo{author}{\bibfnamefont{E.}~\bibnamefont{Artacho}},
  \bibinfo{author}{\bibfnamefont{J.~M.} \bibnamefont{Soler}}, \bibnamefont{and}
  \bibinfo{author}{\bibfnamefont{P.}~\bibnamefont{Ordej{\'o}n}},
  \bibinfo{journal}{Phys.~Rev.~B} \textbf{\bibinfo{volume}{61}},
  \bibinfo{pages}{13639} (\bibinfo{year}{2000}).

\bibitem[{\citenamefont{Di{\'e}guez et~al.}(2001)\citenamefont{Di{\'e}guez,
  Alemany, Rey, Ordej{\'o}n, and Gallego}}]{PRB63-205407}
\bibinfo{author}{\bibfnamefont{O.}~\bibnamefont{Di{\'e}guez}},
  \bibinfo{author}{\bibfnamefont{M.~M.~G.} \bibnamefont{Alemany}},
  \bibinfo{author}{\bibfnamefont{C.}~\bibnamefont{Rey}},
  \bibinfo{author}{\bibfnamefont{P.}~\bibnamefont{Ordej{\'o}n}},
  \bibnamefont{and} \bibinfo{author}{\bibfnamefont{L.~J.}
  \bibnamefont{Gallego}}, \bibinfo{journal}{Phys.~Rev.~B}
  \textbf{\bibinfo{volume}{63}}, \bibinfo{pages}{205407}
  (\bibinfo{year}{2001}).

\bibitem[{\citenamefont{Ordej{\'o}n et~al.}(1995)\citenamefont{Ordej{\'o}n,
  Drabold, Martin, and Grumbach}}]{PRB51-1456}
\bibinfo{author}{\bibfnamefont{P.}~\bibnamefont{Ordej{\'o}n}},
  \bibinfo{author}{\bibfnamefont{D.~A.} \bibnamefont{Drabold}},
  \bibinfo{author}{\bibfnamefont{R.~M.} \bibnamefont{Martin}},
  \bibnamefont{and} \bibinfo{author}{\bibfnamefont{M.~P.}
  \bibnamefont{Grumbach}}, \bibinfo{journal}{Phys.~Rev.~B}
  \textbf{\bibinfo{volume}{51}}, \bibinfo{pages}{1456} (\bibinfo{year}{1995}).

\bibitem[{\citenamefont{Ordej{\'o}n et~al.}(1996)\citenamefont{Ordej{\'o}n,
  Artacho, and Soler}}]{PRB53-10441}
\bibinfo{author}{\bibfnamefont{P.}~\bibnamefont{Ordej{\'o}n}},
  \bibinfo{author}{\bibfnamefont{E.}~\bibnamefont{Artacho}}, \bibnamefont{and}
  \bibinfo{author}{\bibfnamefont{J.~M.} \bibnamefont{Soler}},
  \bibinfo{journal}{Phys.~Rev.~B} \textbf{\bibinfo{volume}{53}},
  \bibinfo{pages}{R10441} (\bibinfo{year}{1996}).

\bibitem[{\citenamefont{Louie et~al.}(1982)\citenamefont{Louie, Froyen, and
  Cohen}}]{PRB26-1738}
\bibinfo{author}{\bibfnamefont{S.~G.} \bibnamefont{Louie}},
  \bibinfo{author}{\bibfnamefont{S.}~\bibnamefont{Froyen}}, \bibnamefont{and}
  \bibinfo{author}{\bibfnamefont{M.~L.} \bibnamefont{Cohen}},
  \bibinfo{journal}{Phys.~Rev.~B} \textbf{\bibinfo{volume}{26}},
  \bibinfo{pages}{1738} (\bibinfo{year}{1982}).

\bibitem[{not({\natexlab{a}})}]{note_on_basis}
\bibinfo{note}{We learned that Di{\'e}guez {\it et~al.} found similar results
  with a basis like ours. Larger extension of basis function is primarily
  essential for precise evaluation of formation energies and less important for
  geometry optimizations (P.~Ordej\'on, private communication).}

\bibitem[{\citenamefont{Moruzzi et~al.}(1986)\citenamefont{Moruzzi, Marcus,
  Schwarz, and Mohn}}]{PRB34-1784}
\bibinfo{author}{\bibfnamefont{V.~L.} \bibnamefont{Moruzzi}},
  \bibinfo{author}{\bibfnamefont{P.~M.} \bibnamefont{Marcus}},
  \bibinfo{author}{\bibfnamefont{K.}~\bibnamefont{Schwarz}}, \bibnamefont{and}
  \bibinfo{author}{\bibfnamefont{P.}~\bibnamefont{Mohn}},
  \bibinfo{journal}{Phys.~Rev.~B} \textbf{\bibinfo{volume}{34}},
  \bibinfo{pages}{1784} (\bibinfo{year}{1986}).

\bibitem[{\citenamefont{Moruzzi}(1986)}]{PRL57-2211}
\bibinfo{author}{\bibfnamefont{V.~L.} \bibnamefont{Moruzzi}},
  \bibinfo{journal}{Phys.~Rev.~Lett.} \textbf{\bibinfo{volume}{57}},
  \bibinfo{pages}{2211} (\bibinfo{year}{1986}).

\bibitem[{\citenamefont{Bagno et~al.}(1989)\citenamefont{Bagno, Jepsen, and
  Gunnarsson}}]{PRB40-1997}
\bibinfo{author}{\bibfnamefont{P.}~\bibnamefont{Bagno}},
  \bibinfo{author}{\bibfnamefont{O.}~\bibnamefont{Jepsen}}, \bibnamefont{and}
  \bibinfo{author}{\bibfnamefont{O.}~\bibnamefont{Gunnarsson}},
  \bibinfo{journal}{Phys.~Rev.~B} \textbf{\bibinfo{volume}{40}},
  \bibinfo{pages}{1997} (\bibinfo{year}{1989}).

\bibitem[{\citenamefont{Leung et~al.}(1991)\citenamefont{Leung, Chan, and
  Harmon}}]{PRB44-2923}
\bibinfo{author}{\bibfnamefont{T.~C.} \bibnamefont{Leung}},
  \bibinfo{author}{\bibfnamefont{C.~T.} \bibnamefont{Chan}}, \bibnamefont{and}
  \bibinfo{author}{\bibfnamefont{B.~N.} \bibnamefont{Harmon}},
  \bibinfo{journal}{Phys.~Rev.~B} \textbf{\bibinfo{volume}{44}},
  \bibinfo{pages}{2923} (\bibinfo{year}{1991}).

\bibitem[{\citenamefont{Herper et~al.}(1999)\citenamefont{Herper, Hoffmann, and
  Entel}}]{PRB60-3839}
\bibinfo{author}{\bibfnamefont{H.~C.} \bibnamefont{Herper}},
  \bibinfo{author}{\bibfnamefont{E.}~\bibnamefont{Hoffmann}}, \bibnamefont{and}
  \bibinfo{author}{\bibfnamefont{P.}~\bibnamefont{Entel}},
  \bibinfo{journal}{Phys.~Rev.~B} \textbf{\bibinfo{volume}{60}},
  \bibinfo{pages}{3839} (\bibinfo{year}{1999}).

\bibitem[{\citenamefont{Blaha et~al.}(2001)\citenamefont{Blaha, Schwarz,
  Madsen, Kvasnicka, and Luitz}}]{wien2k}
\bibinfo{author}{\bibfnamefont{P.}~\bibnamefont{Blaha}},
  \bibinfo{author}{\bibfnamefont{K.}~\bibnamefont{Schwarz}},
  \bibinfo{author}{\bibfnamefont{G.~K.~H.} \bibnamefont{Madsen}},
  \bibinfo{author}{\bibfnamefont{D.}~\bibnamefont{Kvasnicka}},
  \bibnamefont{and} \bibinfo{author}{\bibfnamefont{J.}~\bibnamefont{Luitz}},
  \emph{\bibinfo{title}{{WIEN2k}, {V}ienna {U}niversity of {T}echnology}}
  (\bibinfo{year}{2001}), \bibinfo{note}{improved and updated Unix version of
  the original copyrighted WIEN-code, which was published by P.~Blaha,
  K.~Schwarz, P.~Sorantin, and S.~B.~Trickey, in Comput. Phys. Commun. 59, 339
  (1990)}, \urlprefix\url{http://www.wien2k.at}.

\bibitem[{\citenamefont{Perdew et~al.}(1996)\citenamefont{Perdew, Burke, and
  Ernzerhof}}]{PRL77-3865}
\bibinfo{author}{\bibfnamefont{J.~P.} \bibnamefont{Perdew}},
  \bibinfo{author}{\bibfnamefont{K.}~\bibnamefont{Burke}}, \bibnamefont{and}
  \bibinfo{author}{\bibfnamefont{M.}~\bibnamefont{Ernzerhof}},
  \bibinfo{journal}{Phys.~Rev.~Lett.} \textbf{\bibinfo{volume}{77}},
  \bibinfo{pages}{3865} (\bibinfo{year}{1996}). 
\bibitem[{not({\natexlab{b}})}]{note_on_PAW}
\bibinfo{note}{Of the methods applied so far for the study of small clusters,
  the projector augmented-wave method as used in Ref.~\onlinecite{PRB62-11556}
  has potentially the superior accuracy, as it is an all-electron one and
  allows to enhance the completeness of the basis in a systematic way. However,
  the use of frozen core approximation in actual calculations so far keeps its
  level of accuracy still somehow inferior to, say, FLAPW. The disagreement
  with accurate all-electron results obtained by a Gaussian-type orbital method
  for the Fe$_5$ cluster in Ref.~\onlinecite{APL80-4193} is not yet
  understood.}

\bibitem[{\citenamefont{Freeman and Fu}(1987)}]{JAP61-3356}
\bibinfo{author}{\bibfnamefont{A.~J.} \bibnamefont{Freeman}} \bibnamefont{and}
  \bibinfo{author}{\bibfnamefont{C.~L.} \bibnamefont{Fu}}, \bibinfo{journal}{J.
  Appl. Physics} \textbf{\bibinfo{volume}{61}}, \bibinfo{pages}{3356}
  (\bibinfo{year}{1987}).

\bibitem[{\citenamefont{Spencer et~al.}(2002)\citenamefont{Spencer, Hung,
  Snook, and Yarovsky}}]{SS513-389}
\bibinfo{author}{\bibfnamefont{M.~J.~S.} \bibnamefont{Spencer}},
  \bibinfo{author}{\bibfnamefont{A.}~\bibnamefont{Hung}},
  \bibinfo{author}{\bibfnamefont{I.~K.} \bibnamefont{Snook}}, \bibnamefont{and}
  \bibinfo{author}{\bibfnamefont{I.}~\bibnamefont{Yarovsky}},
  \bibinfo{journal}{Surf.~Sci.} \textbf{\bibinfo{volume}{513}},
  \bibinfo{pages}{389} (\bibinfo{year}{2002}).

\end{thebibliography}



%
%
%
%
%
%
%
%
%
%
%
%

\end{document}